%% file: main.tex
\documentclass[twocolumn,tighten,astrosymb,trackchanges]{aastex631}
\input{affiliations}
\usepackage{CJK}
\usepackage{multirow}
\let\tablenum\relax
\usepackage{siunitx}
\usepackage{appendix}
\usepackage{amsmath,amsthm,amsfonts,amssymb,amscd}
\usepackage{booktabs}
\usepackage{tablefootnote}
\usepackage{apjfonts}
\usepackage{lipsum}
\usepackage{changepage}

\let\ts=\thinspace
\newcommand{\one}{\ts {\sc i}}
\newcommand{\two}{\ts {\sc ii}}
\newcommand{\three}{\ts {\sc iii}}
\newcommand{\four}{\ts {\sc iv}}
\newcommand{\five}{\ts {\sc v}}
\newcommand{\six}{\ts {\sc vi}}
\newcommand{\seven}{\ts {\sc vii}}
\renewcommand\ion[2]{#1\,\,{\sc{\romannumeral #2}}}

\definecolor{maroon}{rgb}{0.760,0.118,0.337}

\def\cm{\mbox{\,cm}}
\def\cm3{\mbox{\,cm$^{-3}$}}

\shorttitle{Spectroscopic time-delay of SN 2025wny}
\shortauthors{Johansson et al.}
\begin{document}

\title{Follow-up of SN~2025wny III: Spectroscopic time-delay measurements of a Strongly Gravitationally Lensed Superluminous Supernova}
  
\correspondingauthor{Joel Johansson}
\email{joeljo@fysik.su.se}

\input{authors_25wny_td}

\begin{abstract}
We present spatially resolved spectra and infer the time-delays between the multiple images of the strongly gravitationally lensed superluminous supernova (SLSN) 2025wny at $z=2.015$. SN~2025wny is the first known spatially resolved strongly lensed SLSN and provides a unique opportunity to measure lensing delays through the temporal evolution of supernova spectra. 
We present a spectroscopic dataset spanning several months, including spatially resolved spectra of images A, B, C, D, and E. We identify and measure the wavelength evolution of spectral features using Gaussian-process modeling. The time delays are inferred by jointly fitting the temporal evolution of the spectral features, yielding $\Delta t_{AB}=-10.3\pm2.3$, $\Delta t_{AC}=0.1\pm3.6$, $\Delta t_{AD}=-65.7\pm3.5$, and $\Delta t_{AE}=3.7\pm8.8$ days (68\% confidence intervals). These are the among most precise time-delay measurements obtained for a lensed supernova to date, whether from spectroscopic or photometric methods. The longest delay ($\Delta t_{AD}$) is particularly well constrained, with a $\sim$5\% precision. Combined with the lens model presented by \citet{Mortsell2026}, the spectroscopic time-delays give a Hubble constant $H_0 = 70.2^{+8.2}_{-6.1}\; {\rm km\,s^{-1}\,Mpc^{-1}}$. Our analysis demonstrates that spectroscopic evolution provides an independent and complementary route to time-delay measurements in lensed supernova systems, avoiding reliance on photometric light curves alone. As future surveys discover larger samples of lensed supernovae, spectroscopic time-delay measurements will provide an important avenue for precision cosmography.
\end{abstract}

\vspace{1.0cm}

\section{Introduction \label{sec:intro}}
Strong gravitational lensing of transient sources provides a direct means of probing both the distribution of matter in foreground lenses and the expansion history of the Universe. As first pointed out by \citet{refsdal_1964b}, photons associated with the same transient event can reach the observer along different paths through a gravitational lens, producing multiple images that appear at different times. The relative arrival times depend on the gravitational potential of the lens and on the time-delay distance, and therefore provide a measurement of the Hubble constant, $H_0$. Time-delay cosmography with strongly lensed supernovae (glSNe) is particularly attractive because it provides a route to $H_0$ that is independent of the traditional distance ladder and early-Universe constraints. In addition, supernovae offer well-defined temporal evolution on timescales of weeks to months, fade after the explosion, and (in the case of Type~Ia supernovae) provide absolute magnification information through their standardizable luminosities. Beyond cosmography, the lensing magnification acts as a "gravitational telescope", enabling detailed studies of intrinsically faint or very distant stellar explosions that would otherwise be inaccessible to high signal-to-noise photometric and spectroscopic follow-up \citep[see e.g.][for recent reviews]{suyu2024SSRv..220...13S,Goobar2025RSPTA.38340123G}.

The observational realization of Refsdal's proposal came five decades later with the discovery of the multiply imaged SN~Refsdal \citep{kelly_2015,Kelly_2016}. Since then, systematic searches of strongly lensed galaxy clusters with the Hubble Space Telescope (HST) and the James Webb Space Telescope (JWST) have resulted in a growing sample of cluster-lensed transients spanning a range of supernova types and redshifts \citep[e.g.,][]{rodney_2021,chen_2022,pierel_2024}. A particularly important example is the triply imaged Type~Ia SN~H0pe \citep{Frye2024ApJ...961..171F}, whose relative time-delays were constrained independently from its photometric evolution and from the spectroscopic phases of the three images \citep{Pierel2024ApJ...967...50P,Chen2024ApJ...970..102C}, and subsequently combined with cluster lens models to infer $H_0$ \citep{Pascale2025ApJ...979...13P,Grayling2026MNRAS.548ag340G}. 

In parallel, wide-field ground-based time-domain surveys provide a complementary discovery channel for supernovae strongly lensed by individual galaxies. The first spatially resolved multiply imaged Type~Ia supernova discovered in such a system, iPTF16geu \citep{goobar_2017}, was followed by SN~Zwicky \citep{goobar_2023}. The compact image configurations of these systems produced delays of only order days or less \citep{Dhawan2020MNRAS.491.2639D,Johansson2021MNRAS.502..510J}, limiting their usefulness for precision time-delay cosmography, but demonstrated that galaxy-scale glSNe can be identified efficiently in untargeted wide-field surveys. More recently, the ground-based sample has begun to expand: the Type~II SN~2025mkn \citep{Lemon2026ApJ..1003L..47L}, the Type~Ia SN~2026ngr \citep{Johansson2026TNSAN.156....1J} and the Type I SLSN 2025wny. These discoveries illustrate the rapidly increasing diversity of glSNe found by wide-field surveys and foreshadow the substantially larger samples expected from the Vera C. Rubin Observatory's Legacy Survey of Space and Time (LSST).

\subsection{SN 2025wny}
SN~2025wny represents a particularly favorable addition to the emerging population of glSNe. Discovered by ZTF and independently reported by GOTO, SN~2025wny was spectroscopically classified as a hydrogen-poor superluminous supernova (SLSN-I) at $z\simeq2.01$ \citep{Johansson2025ApJ...995L..17J,Taubenberger2026A&A...710A.365T}. High-resolution imaging resolved five images of the SN around a pair of massive foreground galaxies at $z\simeq0.375$. 

SN~2025wny is the first spatially resolved strongly lensed SLSN and, in contrast to previously discovered galaxy-scale glSNe, combines relatively wide image separations with delays extending to several tens of days. Its high intrinsic luminosity and substantial lensing magnification have enabled an unusually extensive spectroscopic follow-up campaign, with spatially resolved spectra of multiple images obtained repeatedly over more than 200 observer-frame days. The system therefore provides both an opportunity for time-delay cosmography and a unique view of the spectroscopic evolution of a high-redshift SLSN.

In this work we present spatially resolved spectroscopy of the five images A--E of SN~2025wny obtained over several months with a heterogeneous set of ground-based spectrographs. We identify broad emission and absorption features that can be followed consistently across instruments and epochs, measure their extrema using Gaussian-process regression, and jointly fit their temporal evolution to infer the relative time-delays. 
We compare the spectroscopic time-delays with the independent photometric measurements of \citet{Townsend2026} and combine them with the lens models of \citet{Mortsell2026} to explore the resulting constraints on $H_0$. Accompanying papers present the space-based observations of the system \citep{Goobar2026}, the detailed spectrophotometric evolution and physical interpretation of the SLSN \citep{Li2026}, the host galaxy properties \citep{Qin2026}, and the expected detection rate of lensed SLSNe-I in the Zwicky Transient Facility survey \citep{Hjortlund2026}.

\section{Observations \label{sec:obs}}
\begin{figure*}
    \centering
    \includegraphics[width=0.99\textwidth]{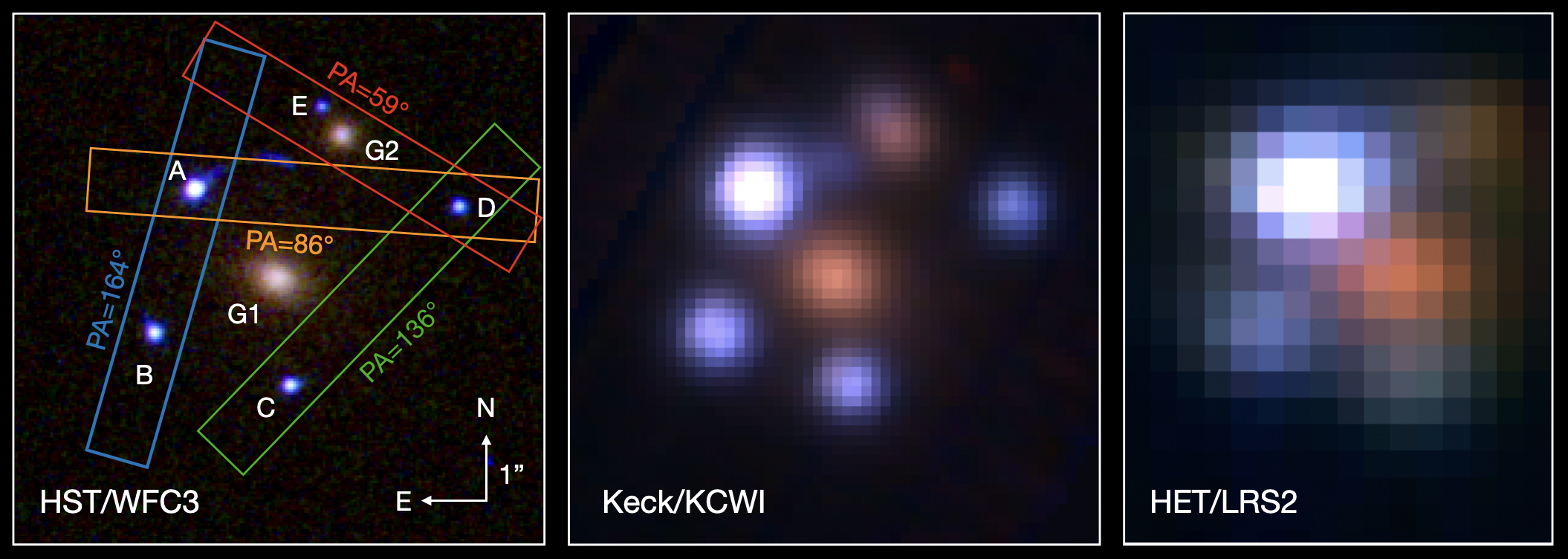}
    \caption{The left panel shows the slit orientations used for the long-slit observations, overlaid on a HST color composite image from \citet{Goobar2026}. Middle and right panels show ground-based Integral Field Unit (IFU) data as color composite images of summed slices centered at 4000/5000/6000~\AA, from Keck/KCWI (middle) and HET/LRS2 (right) observations.
    }
    \label{fig:slits_ifu}
\end{figure*}

Early spectra of the brightest image (A) were used in \citet{Johansson2025ApJ...995L..17J} and \citet{Taubenberger2026A&A...710A.365T} to establish the SN type and redshift of SN\,2025wny. 

Here we present additional, spatially resolved spectra of images A, B, C, D and E of SN\,2025wny. In total, we  20 spectra of image A, 15 spectra of image B, 5 spectra of image C, 5 spectra of image D and 2 spectra of image E, observed over the course of more than 200 days. The log of spectra used in this analysis is presented in \autoref{tab:obs}.

\subsection{Spectroscopy \label{sec:obs_spec}}
For long-slit observations, different position angles (PA, measured counterclockwise from celestial north) were used.
\autoref{fig:slits_ifu} shows the slit positions, covering images A and B ($PA = 164\degr$, separated by 2\farcs2), images C and D ($PA=136\degr$, separated by 3\farcs7), images D and E ($PA = 59\degr$, separated by 2\farcs5) or images A and D ($PA=86\degr$, separated by 3\farcs9). 

The Low-Resolution Imaging Spectrometer \citep[LRIS;][]{Oke95} on the Keck I telescope was used to obtain four epochs of spectroscopy, with a 1\farcs0 wide slit using the B400/3400 grism and R400/8500 grating at several different position angles. Spectra were reduced and extracted using \texttt{LPipe} \citep{2019PASP..131h4503P} and \texttt{PypeIt} \citep{prochaska_pypeit2020}. 

We obtained six epochs of spectroscopy with the FOcal Reducer/low dispersion Spectrograph 2 \citep[FORS2;][]{Appenzeller1998Msngr..94....1A} at the ESO 8.2\,m Very Large Telescope (VLT) at Paranal Observatory (Chile, DDT programme 2116.A-5022(A), PI: Dhawan). We used the 300V grism and a 1\farcs0 slit covering images A and B, providing an instantaneous wavelength coverage from 3800--9600~\AA\,at a spectral resolution of $R\sim440$ based on the FORS2 documentation. The spectra were reduced with \texttt{PypeIt} \citep{prochaska_pypeit2020}.

Two epochs of spectroscopy were obtained with the GMOS spectrograph on the Gemini North telescope. The first epoch was observed in long-slit mode covering images A and B (program GN-2025B-DD-106, PI: Dhawan) using a 1\farcs0 slit and the R400 grating. The second epoch used the Multi-Object Spectroscopy (MOS) mode (program GN-2025B-FT-108, PI: Qin). Several slitlets were used covering images A, B and D and the host galaxy arc \citep[see details in][]{Qin2026}. 

One epoch of spectroscopy was obtained with the upgraded Optical System for Imaging and low Resolution Integrated Spectroscopy \citep[OSIRIS$+$;][]{Cepa2000SPIE.4008..623C} instrument at the 10.4 m Gran Telescopio Canarias  (GTC) at the Roque de los Muchachos Observatory (Spain) under DDT programme: GTC2025-284 (PI: Prada). We utilised the R300B grating covering the wavelength range from 3800 to 10000\,\AA. The 1\farcs0 wide slit covered images A and D, at a spectral resolution of $R\sim216$ according to instrument documentation.

One spectrum of image A was obtained with the Binospec spectrograph \citep{Fabricant2019PASP..131g5004F} on the 6.5-m MMT using a 1\farcs0\ slit and the 270 lines mm$^-1$ grating. The wavelength coverage is 3900 to 9200\,\AA, with a spectral resolution $R\sim1340$. The spectrum was reduced with \texttt{PypeIt} \citep{prochaska_pypeit2020}.

Three IFU spectroscopic observations are used in this analysis \citep[not including the two JWST/NIRSPEC IFU observations presented in][]{Goobar2026,Li2026}.

Two observations were obtained using the \emph{Keck Cosmic Web Imager} (KCWI) mounted on the Nasmyth platform of the Keck II telescope. 
We employed the small slicer configuration, providing a field of view of $\sim$8" $\times$ 20" covering all five SN images and the two lens galaxies. The spectral resolutions are $R \sim 3600$ and $R \sim 2000$ for the blue and red arm, respectively. Further details on the observation and reduction procedures can be found in \citet{Qin2026}.

In addition, one epoch of observations was conducted with the Low Resolution Spectrograph (LRS2; \citealt{Chonis2014,Chonis2016}) integral field unit (IFU) in its blue channel ($\sim$3650 to 6950 \AA) on the Hobby–Eberly Telescope (HET; \citealt{1998SPIE.3352...34R, 2021AJ....162..298H}) under program M25-3-005 (PI: D. Gruen). Observations were scheduled using the HET queue-scheduling system \citep{2007PASP..119..556S}. 
The IFU provides contiguous spatial sampling over a 12" × 6" field of view, covering images A, B, and C. Data were processed using \texttt{Panacea}\footnote{\url{https://github.com/grzeimann/Panacea}} and \texttt{LRS2Multi}.\footnote{\url{https://github.com/grzeimann/LRS2Multi}}

\begin{table*}
    \centering
    \begin{tabular}{lllllllll}
        \hline
        \textbf{Images} & \textbf{Date} & \textbf{MJD} & \textbf{$t_{\rm obs}-t_0$} & \textbf{Telescope / Instrument} & \textbf{Wavelength range (\AA)}  & \textbf{Airmass} & \textbf{Seeing (")} & Source \\ 
        \hline
        \hline
        A & 2025-09-04 & 60922.19 & 11.7 & NOT / ALFOSC & 3300 -- 9000 & 2.17 & 1.0 & J25 \\ 
        A & 2025-09-24 & 60942.51 & 32.0 & P200 / NGPS & 5600 -- 10000 & 1.16 & -- & L26 \\ %
        A & 2025-09-26 & 60944.16 & 33.7 & NOT / ALFOSC & 3300 -- 9000 & 1.64 & 0.9 & J25 \\ 
        A & 2025-10-11 & 60959.14 & 48.6 & NOT / ALFOSC & 3500 -- 9700 & -- & -- & T26 \\ 
        A & 2025-10-15 & 60963.55 & 53.0 & UH88 / SNIFS & 3500 -- 9700 & -- & -- & T26 \\ 
        A & 2025-10-19 & 60967.15 & 56.7 & NOT / ALFOSC & 3500 -- 5400 & -- & -- & T26 \\ 
        \hline
        ABCDE  & 2025-10-22 & 60970.56 & 60.1 & Keck I / LRIS & 3200 -- 10200 & 1.17 & -- & J25/J26 \\ 
        ABC & 2025-10-31 & 60979.37 & 68.9 & HET / LRS2 (IFU) & 3700 -- 7000 & 1.22 & 1.4 & J26 \\ 
        AB & 2025-11-16 & 60995.50  & 85.0 & Gemini-N / GMOS & 3700 -- 7000 & 1.29 & -- & J26 \\ 
        ABCDE  & 2025-11-25 & 61004.52 & 94.0  & Keck II / KCWI (IFU) & 3500 -- 10200 & 1.06 & 0.7 & J26/Q26 \\ 
        AD  & 2025-11-28 & 61008.07 & 97.6 & GTC / OSIRIS$+$  & 3000 -- 9800 & 1.06 & 1.1 & J26 \\ 
        A & 2025-11-29 & 61008.35 & 97.8 & MMT / BINOSPEC & 3900 -- 9200 & 1.09 & -- & J26 \\ 
        AB & 2025-12-15 & 61024.27 & 113.8 & VLT / FORS2 & 3400 -- 9600 & 2.19 & 0.8 & J26 \\ 
        AB & 2025-12-18 & 61027.40 & 116.9 & Keck I / LRIS & 3200 -- 10200 & 1.21 & 0.8 & J26 \\ 
        AB & 2025-12-21 & 61030.34 & 119.8 & Keck I / LRIS & 3200 -- 10200 & 1.44 & -- & J26 \\ 
        AB & 2025-12-24 & 61033.26 & 122.8 & VLT / FORS2 & 3400 -- 9600 & 2.20 & 0.7 & J26 \\
        ABCD & 2026-01-11 & 61051.00 & 140.5 & Keck II / KCWI & 3500 -- 9000 & -- & 1.0  & J26/Q26 \\
        AB  & 2026-01-18 & 61058.16 & 147.7 & VLT / FORS2 & 3400 -- 9600 & 2.20 & 1.0 & J26 \\ 
        ABD & 2026-01-20 & 61060.36 & 149.9 & Gemini-N / GMOS (MOS) & 3500 -- 7600 & 1.10 & -- & J26/Q26 \\ 
        AB & 2026-02-09 & 61080.08 & 169.6 & VLT / FORS2 & 3400 -- 9600 & 2.24 & 1.5 & J26\\ 
        AB & 2026-02-18 & 61090.26 & 179.8 & Keck I / LRIS & 3200 -- 10200 & 1.12 & -- & J26\\ 
        AB & 2026-03-15 & 61114.01 & 203.5 & VLT / FORS2 & 3400 -- 9600 & 2.19 & 0.7 & J26\\ 
        AB & 2026-04-08 & 61138.01 & 227.5 & VLT / FORS2 & 3400 -- 9600 & 2.53 & 1.1 & J26\\ 
             \hline
    \end{tabular}
    \caption{Spectroscopic data used in this paper. The epoch of observation $t_{\rm obs}-t_0$ is in observer frame days from MJD=60910.50. Sources of the spectra are labeled J25: \citet{Johansson2025ApJ...995L..17J}, T26: \citet{Taubenberger2026A&A...710A.365T} and J26: This work, including accompanying papers by \citep[L26;][]{Li2026} and \citep[Q26;][]{Qin2026}.
    }
    \label{tab:obs}
\end{table*}
\subsection{Spectral extractions}\label{sec:spectralextractions}
\begin{figure*}[htp]
    \centering
    \includegraphics[width=0.9\textwidth]{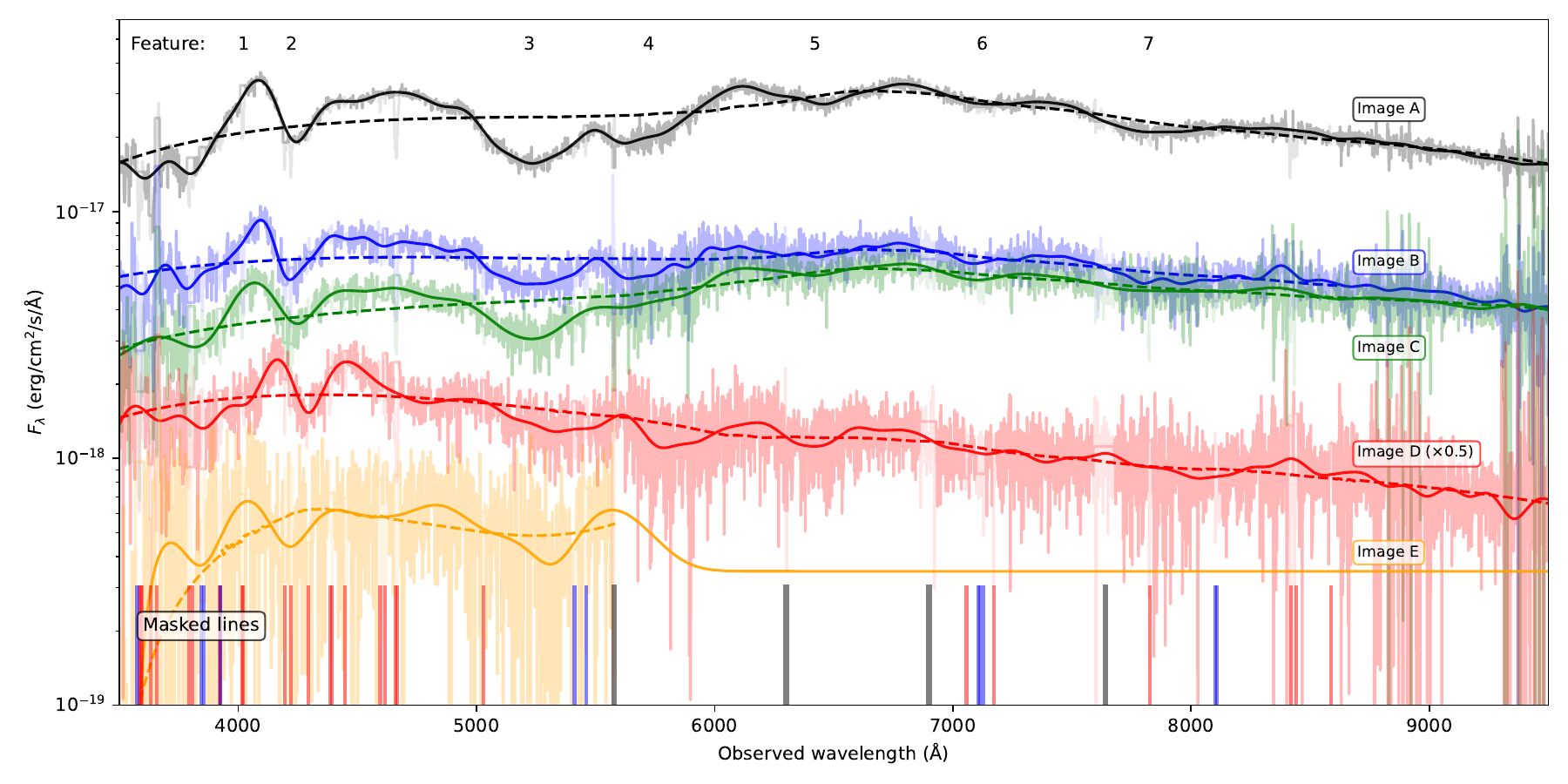}
    \caption{Individual spectra of images A, B, C, D and E from Keck/LRIS on 2025 Oct. 22. The location of a sharp emission feature (feature~1) and six broad absorption features (2-7) are labeled at the top. The thin vertical lines at the bottom indicate the location of narrow absorption features from the host (red), lens (blue) and telluric absorption (black) that were masked when fitting the broad spectral features. The dashed curves show the fitted continuum levels, which are used to normalize the spectra. Note that the spectrum of Image E was recovered after subtracting out contaminating flux from lens galaxy G2 (see details in \autoref{sec:galaxies}).
    }
    \label{fig:keck_spec}
\end{figure*}
For the VLT/FORS2 (covering images A and B) and GTC/OSIRIS+ (covering images A and D) observations, we extracted one-dimensional spectra from the sky-subtracted two-dimensional frame using a custom implementation of optimal extraction. This follows the formalism of \citet{Horne1986a}, extended to partially overlapping traces after \citet{Marsh1989PASP..101.1032M}.

Images A and B are sometimes blended, so we modeled the two traces jointly. We first binned the two-dimensional spectrum along the dispersion direction and computed the mean cross-dispersion profile in each bin, sigma-clipping to reject cosmic rays and other outliers. In each bin we fitted this profile as the sum of two Gaussians of common width,
\begin{equation}
M(y) = A_1 \exp\!\left[-\frac{(y-\mu_1)^2}{2\sigma^2}\right]
     + A_2 \exp\!\left[-\frac{(y-\mu_2)^2}{2\sigma^2}\right] ,
\end{equation}
where $y$ is the cross-dispersion coordinate, $\mu_1$ and $\mu_2$ are the spatial centroids of the two traces, $\sigma$ is their shared width, and $A_1$ and $A_2$ are the component amplitudes. This fit is used only to measure $\mu_1$, $\mu_2$ and $\sigma$. The amplitudes anchor the fit to the observed profile but are not flux measurements, and they do not enter the extraction step described below.

An initial pass with all parameters free gave the trace separation. We fixed the separation to the median value from the well-constrained bins and refitted with four free parameters, $(A_1,\mu_1,\sigma,A_2)$. We then fitted the resulting centroids and widths as functions of position along the dispersion axis, $x$, using Chebyshev polynomials of order 2 (for $\sigma$) or 3 (for the center of the primary trace). This gives smooth, full-resolution models $\mu_1(x)$, $\mu_2(x)$ and $\sigma(x)$, and removes column-to-column noise in the profile model while preserving the slowly varying instrumental point spread function.

For the flux extraction itself, we did not reuse $M(y)$ directly. Instead, at each $x$ we built a separate, unit-area Gaussian for each trace from the smoothed centroid and width alone,
\begin{equation}
P_i(y) = \frac{1}{\sigma(x)\sqrt{2\pi}} \exp\!\left[-\frac{(y-\mu_i(x))^2}{2\sigma(x)^2}\right], \qquad i = 1, 2 .
\end{equation}
Unlike the two terms of $M(y)$, $P_1(y)$ and $P_2(y)$ each integrate to unity on their own and carry no amplitude information: only the trace position and width matter. We measured the fluxes on the unbinned frame by solving, at each wavelength, for the vector $\mathbf{f}=(f_1,f_2)$ of trace fluxes that best reproduces the observed sky-subtracted spatial profile $D(y)$ as a combination of these two templates,
\begin{equation}
D(y) = f_1 P_1(y) + f_2 P_2(y) .
\end{equation}
We used inverse-variance weighting throughout and obtained the solution from the $2\times2$ normal equations. Fitting the two profiles simultaneously accounts for the overlap and removes cross-talk between the two traces to first order.

For the IFU observations, we extracted the spectra of the individual SN images and foreground lens galaxies using a scene-modeling approach. First, a white-light image was constructed by stacking the full IFU cube (see middle and right panels of \autoref{fig:slits_ifu}) which was used to refine the astrometric alignment, fit Sérsic profiles for the lens galaxies and PSF components for the supernova images. The model was then applied to each wavelength slice of the IFU cube, keeping the galaxy shapes, source positions, and PSF parameters fixed to the white-light solution, fitting only the amplitudes of the individual galaxy and supernova components, together with a constant background term. 

Image E lies in close projection to one of the lens galaxies (0.5" from the center of G2) and is significantly contaminated by galaxy light. We recover the blue portion ($<$5500 \AA) of image E at two epochs by subtracting the lens-galaxy spectrum of G2 reconstructed from the Keck/KCWI IFU scene-model fit. This subtraction removes both the stellar continuum and narrow absorption features associated with the foreground galaxy while preserving the broad supernova features used in the subsequent analysis.

In Appendix \ref{sec:galaxies}, we analyze the extracted lens-galaxy spectra of G1 and G2 with the Penalized Pixel-Fitting \texttt{pPXF} code \citep{Cappellari2023}, fitting the stellar continua with simple stellar population templates while simultaneously optimizing the galaxy redshifts and stellar velocity dispersions.

\section{Spectral Analysis \label{sec:spec_analysis}}
Spectroscopic phase information has previously been used to measure lensing delays for the multiply imaged Type Ia SNe iPTF16geu and SN H0pe, by fitting spectral templates to the individual images to determine their relative ages \citep{Johansson2021MNRAS.502..510J,Chen2024ApJ...970..102C}. Template-fitting approaches, however, rely on densely time-sampled, well-calibrated spectral template libraries that presently exist only for a handful of well-studied SN classes (mainly SNe Ia). No comparable template set exists for SLSNe-I, whose rest-frame UV spectroscopic evolution remains only sparsely characterized \citep{Yan2018ApJ...858...91Y}. We therefore develop a template-independent framework that instead tracks the temporal evolution of individual emission and absorption features directly, similar to what was proposed for simulated lensed Type II SNe \citep{Bayer2021A&A...653A..29B}. 

We select seven broad spectral features that can be identified and followed across multiple epochs for the different lensed images: a sharp emission feature near 4150~\AA\, (feature 1) and six broad absorption features (labeled as 2-7, see \autoref{fig:keck_spec}). In particular, we favored broad features with well-defined minima or maxima that remain measurable despite differences in spectral resolution, wavelength coverage, and signal-to-noise ratio between epochs.

Each spectrum was analyzed independently using a semi-automated procedure. Prior to measuring individual broad line positions, narrow host and lens galaxy absorption lines, together with strong telluric absorption bands, were masked (see vertical lines at the bottom of \autoref{fig:keck_spec}). Our highest resolution spectra reveal numerous narrow absorption lines matching strong interstellar features at $z$=2.011 and 2.014, close to the SN host redshift $z=2.0151$ \citep[see][for details]{Johansson2025ApJ...995L..17J,Goobar2026,Qin2026}. 
Masking out these lines is most important for Features 1 and 2, which e.g. move in and out of the \ion{Si}{4} $\lambda\lambda$1394,1403 lines.

Then, a pseudo-continuum (dashed lines in \autoref{fig:keck_spec}) was estimated using a broad smoothing window and a third-order polynomial, using the pre-processing utilities in deepSIP \citep{Stahl_2020MNRAS.496.3553S}, and each spectrum was normalized by dividing by this pseudo-continuum. This step is not crucial, but helps to homogenize the spectra, since perfect absolute flux calibration is not guaranteed given the different observing conditions (some spectra observed far from parallactic angle, at high airmass, with/without atmospheric dispersion correction,  etc.). Although the normalization has only a minor effect on the measured extrema of the strongest features, it improves the robustness of the measurements for shallower features (e.g., Features 4 and 6), particularly when they lie on steep continua

\subsection{Spectral Feature Measurements}\label{sec:features_measured}
The broad features in SLSN spectra are generally blends of multiple transitions, and their detailed composition may evolve with time \citep[see][]{Li2026}. For the purpose of measuring time-delays, we do not attempt to associate each feature with a particular ion or transition, but instead treat the measured minima and maxima as empirical tracers of the spectral evolution.

Rather than assuming an analytic line profile, each feature was modeled non-parametrically using GP regression.  We adopted a Matérn covariance kernel, whose characteristic length scale was optimized during the GP fit to reproduce the local structure of the line profile while suppressing high-frequency noise. Separate GP fits were performed for absorption minima and emission maxima using identical procedures.
Each spectral feature was measured iteratively. We first identified the minimum (or maximum) within a predefined wavelength interval, then recentered the fitting window on this initial estimate and refitted the feature. The final windows, typically 300--600~\AA\ wide, were chosen to encompass the broad feature while avoiding neighboring features.

\begin{figure}
    \centering
    \includegraphics[width=\columnwidth]{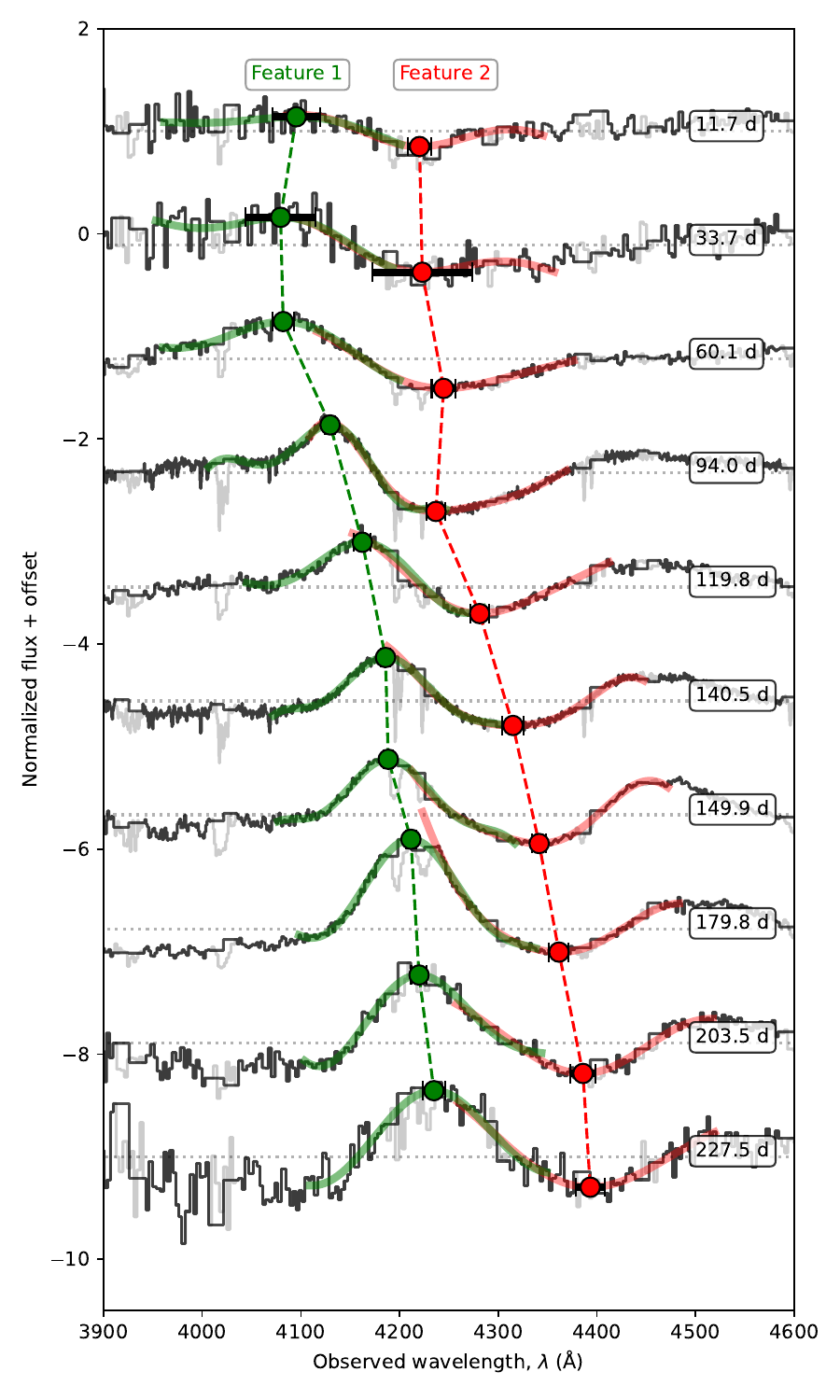}
    \caption{Example of GP fits to the emission feature near 4150~\AA\, (feature 1, green lines and symbols) and absorption feature near 4300~\AA\, (feature 2, red lines and symbols) for a subset of spectra of Image A.}
    \label{fig:gpfit}
\end{figure}

The line position was determined from the extremum of the GP model on a finely sampled wavelength grid. Uncertainties were estimated from Monte Carlo realizations of the GP posterior, with the median extremum adopted as the line position and the 16th--84th percentiles as its uncertainty, naturally accounting for both measurement noise and uncertainty in the reconstructed line profile. \autoref{fig:gpfit} shows the GP fits to the emission feature near 4150\,\AA\,
(feature 1, green lines and symbols) and absorption feature near
4300\,\AA\,(feature 2, red lines and symbols) for a subset of spectra
of Image A.

This non-parametric approach avoids imposing a Gaussian, polynomial, or other analytic line shape, making it well suited for the broad, asymmetric, and often blended features observed in supernova spectra. The resulting measurements (see \autoref{tab:features}) provide a homogeneous set of feature wavelengths and uncertainties that form the input to the joint time-delay analysis described in the following section.

\subsection{Joint time-delay inference from spectral features}
We infer relative time-delays between multiple lensed images A, B, C, D and E by exploiting the temporal evolution of the seven spectral features described and measured in \autoref{sec:features_measured} and \autoref{tab:features}.

We assume that all images probe a common intrinsic spectral evolution described by a latent function $f(t)$, but are offset in time due to lensing delays. Let $\lambda_{i,j}(t)$ denote the measured wavelength for feature $i$ $\left(i \in \{1,\dots,7\}\right)$ for image $j$ $\left(j \in \{A,\dots,E\}\right)$ at observed epoch $t$. The model can be written as
\begin{equation}
    \lambda_{i,j}(t) \approx f_i(t - \tau_j), 
\end{equation}
where $\tau_j$ is the time-delay of image $j$ relative to a chosen reference image (A, i.e. $\tau_A \equiv0$). To infer the delays, we adopt a forward-modeling approach in which:
\begin{enumerate}  
    \item  Each trial set of delay parameters ($\tau_{B}$, $\tau_{C}$, $\tau_{D}$ and $\tau_{E}$) is applied to the corresponding time series.
    \item  The shifted datasets, together with measurements for Image A, are combined into a single pooled dataset.
    \item  A weighted smoothing spline or a GP with fixed kernel hyperparameters is fit to this pooled dataset, representing the common intrinsic evolution $f(t)$ .
    \item  The likelihood is computed from the residuals of all datasets with respect to the spline or GP predictive mean.
\end{enumerate}

\begin{figure*}
    \centering
    \includegraphics[width=0.70
    \textwidth]{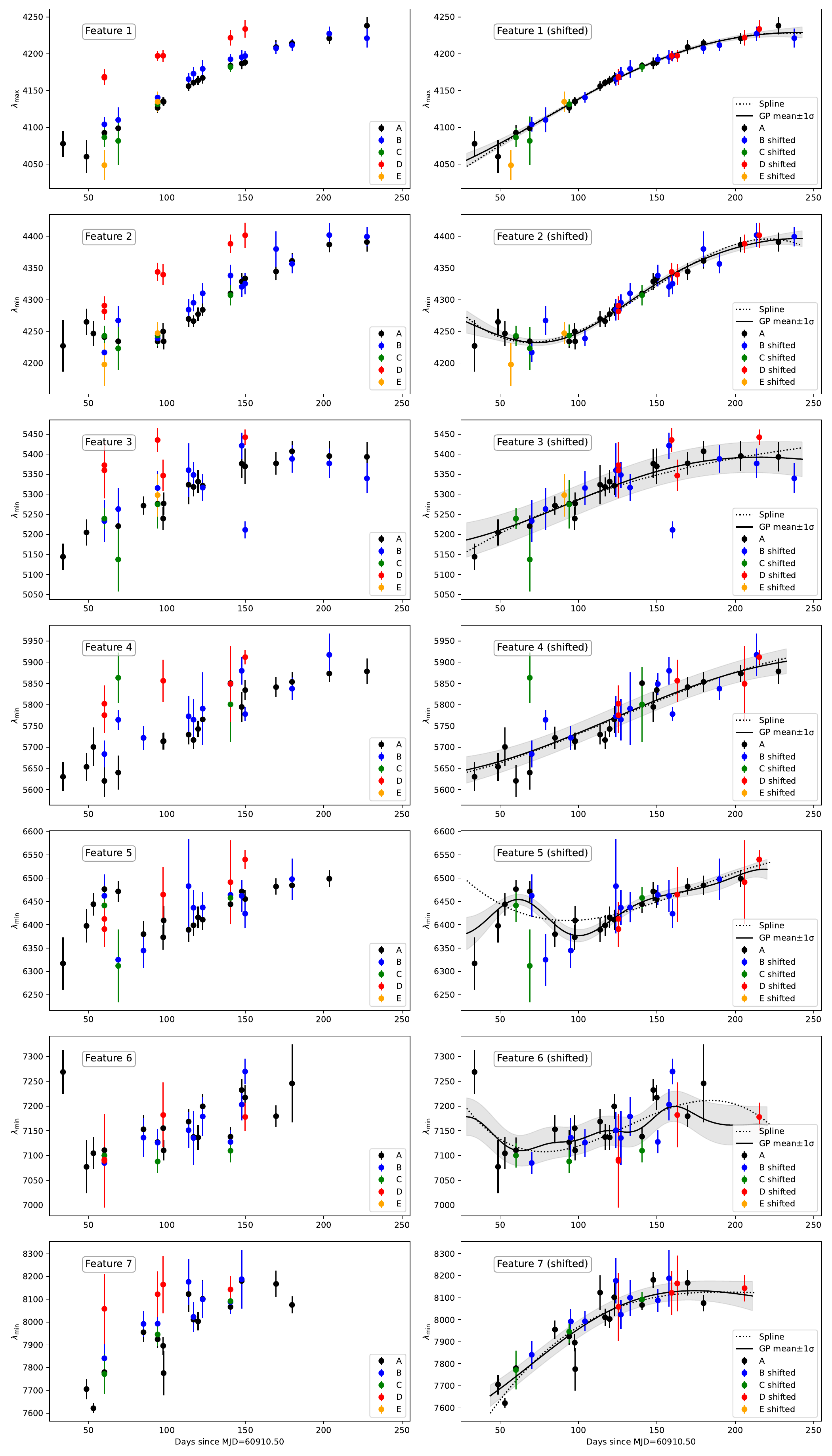}
    \caption{\emph{Left panels:} Measured wavelength maxima/minima for features 1-7 at all observed epochs for images A, B, C, D and E. \emph{Right panels:} Time evolution of the fitted maximum/minima of the  spectral features after the best-fit time-delays are applied. Solid and dotted lines show the fitted GP and spline functions representing the common intrinsic evolution, respectively.
    }
    \label{fig:features}
\end{figure*}

\autoref{fig:features} (left panels) shows the measured wavelength maxima/minima for features 1 -- 7 at all observed epochs for images A, B, C, D and E. The right panels show the time evolution of the fitted maximum/minima of the spectral features after the best-fit time-delays are applied. Solid and dotted lines show the fitted GP and spline functions representing the common intrinsic evolution, respectively.

Several trends are apparent in the joint fit (\autoref{fig:features}). Features 1–3 are the sharpest and most cleanly defined in individual spectra and show the tightest constraints on their common evolution once shifted by the best-fit delays. The redder Features 4–7, being shallower and broader, show larger scatter in their individual epoch measurements, but remain broadly consistent with the fitted common evolution after the joint fit. All seven features shift toward redder wavelengths with time, with net displacements of order $\sim$150 -- 400\,\AA\, over the $\sim$200 day observed baseline (at an average rate of $\sim1 - 2$\,\AA\, per observer-frame day).
Although not enforced in the fit, most features increase in wavelength monotonically. Features 5 and 6 are the exception, showing modest structure (a flattening or brief reversal) at early phases. This could reflect either a genuine physical origin (e.g., a shift in the dominant transition contributing to the blend as the SN evolves) or a measurement systematic (e.g., residual contamination from a masked narrow line near the feature). Distinguishing between these is not important for the time-delay measurement itself since our method is agnostic to which species or blend of species is responsible for a given feature. The joint fit recovers the relative delay as long as the same feature is consistently (mis-)identified across all five lensed images.

We derive posterior uncertainties on the four shared delays by sampling the joint posterior with an affine-invariant MCMC ensemble sampler \citep[\texttt{emcee};][]{foreman-mackey_emcee_2013}, assuming flat priors over the grid-search bounds. As a cross-check, we also perform a parametric bootstrap, in which each measured feature extremum is perturbed within its reported uncertainty and the full joint time-delay fit is repeated. The two approaches yield consistent results, providing an additional check on the robustness of the inferred uncertainties.

Throughout this work, we adopt the convention $\Delta t_{AX} \equiv t_X - t_A = \tau_X$ (such that negative values indicate images arriving before Image A). 

\begin{figure*}[htp]
    \centering
    \includegraphics[width=\textwidth]{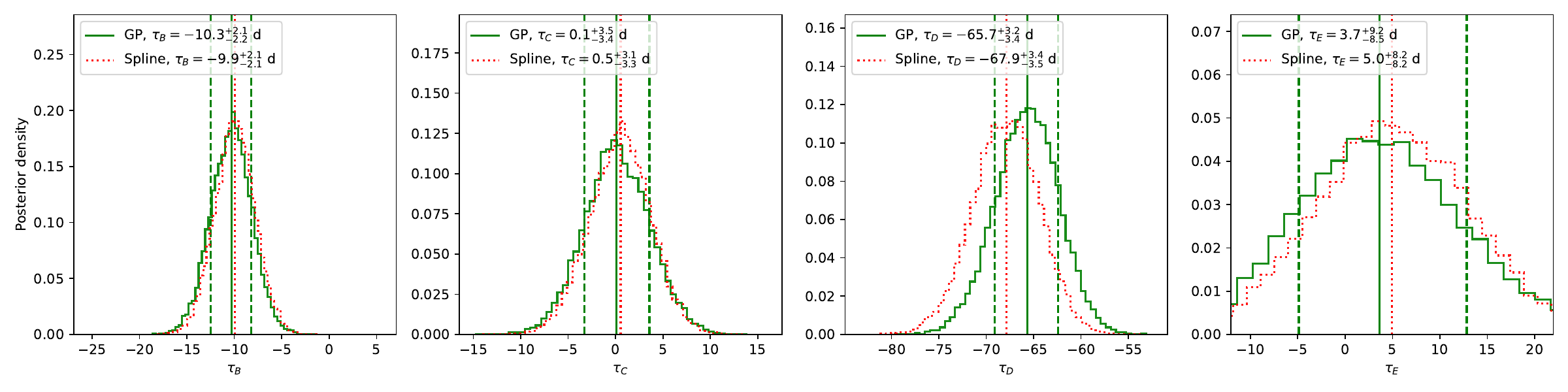}
    \caption{Posterior distributions of the time-delay parameters $\tau_{B}$, $\tau_{C}$, $\tau_{D}$ and $\tau_{E}$ using all spectral features (1 to 7). Green solid lines show the fiducial result using the GP based interpolating function while red dotted lines show results using the spline interpolator.}\label{fig:posteriors2}
\end{figure*}

The inferred time-delays relative to the reference Image A are listed in \autoref{tab:delays} for the two choices of interpolating function, a smoothing spline and a GP model. The two approaches yield consistent delays (green and red curves in \autoref{fig:posteriors2}), indicating that the results are not strongly dependent on the adopted representation of the intrinsic spectral evolution. We adopt the GP results as our fiducial measurements. The GP provides a flexible, non-parametric description of the latent evolution of each spectral feature without specifying the location of spline knots or imposing a particular functional form. Moreover, its covariance kernel provides a natural scale over which measurements at neighboring epochs are correlated, making it well suited to the irregular temporal sampling and heterogeneous uncertainties of our spectroscopic data. We therefore regard the GP as the more natural representation of the unknown underlying feature evolution, while using the independent spline fits as a test of the sensitivity to this choice. Our fiducial delays are $\Delta t_{AB}=-10.3\pm2.3$, $\Delta t_{AC}=0.1\pm3.6$, $\Delta t_{AD}=-65.7\pm3.5$, and $\Delta t_{AE}=3.7\pm8.8$ days, where the uncertainties correspond to the 68\% confidence intervals derived from the posterior samples.
\begin{table}
    \centering
    \begin{tabular}{lcccc}
        \hline
        \textbf{Method} & $\Delta t_{AB}$ & $\Delta t_{AC}$ & $\Delta t_{AD}$ & $\Delta t_{AE}$  \\ 
        \hline
        \hline
        GP      & $-10.3^{+2.1}_{-2.2}$ & $+0.1^{+3.5}_{-3.4}$ & $-65.7^{+3.2}_{-3.4}$ & $+3.7^{+9.2}_{-8.5}$ \\
        spline  & $-9.9^{+2.1}_{-2.1}$  & $+0.5^{+3.1}_{-3.3}$ & $-67.9^{+3.4}_{-3.5}$ & $+5.0^{+8.2}_{-8.2}$ \\
        \hline
    \end{tabular}
    \caption{Inferred time-delays (in observer-frame days) obtained from the joint fit to all of the measured spectral features (1–7), using the GP and spline models.
    }
    \label{tab:delays}
\end{table}

As seen in \autoref{fig:posteriors2}, the posterior distributions are unimodal and approximately Gaussian. The posterior distributions of the four relative time delays exhibit only weak covariance, with all pairwise correlation coefficients satisfying $|\rho|<0.15$. We therefore find no strong degeneracies between the individual time-delay measurements in the joint fit.

\section{Discussion}\label{sec:discussion}
The spectroscopic time-delays of SN~2025wny are measured with precisions ranging from $\sim2$ days for $\Delta t_{AB}$ to $\sim3.5$ days for $\Delta t_{AC}$ and $\Delta t_{AD}$. These are among the most precise time-delay measurements obtained for a lensed supernova to date, whether from spectroscopic or photometric methods. The precision is enabled by the combination of a large number of individual feature measurements across multiple images and epochs (hence the larger uncertainty for $\Delta t_{AE}$), together with the long temporal baseline over which the spectral evolution is traced.

In relative terms, the longest delay, $\Delta t_{AD}$, is the most precisely constrained, with a fractional uncertainty of only $\sim5\%$. The shorter delays have correspondingly larger fractional uncertainties despite comparable absolute precision, with $\Delta t_{AB}$ measured to $\sim20\%$ precision.

The single-epoch spectroscopic time delays inferred for iPTF16geu and SN~H0pe \citep{Johansson2021MNRAS.502..510J,Chen2024ApJ...970..102C} relied on matching the rest-frame spectral phases of the individual images, which can be achieved to a precision of $\sim$2 -- 4 days. However, when converted to observer-frame time delays, this uncertainty is inevitably amplified by a factor of $(1+z_{\rm SN})$.

Several potential sources of systematic bias in the spectroscopic measurements are worth noting for future work. Host- and lens-galaxy light contamination can, in principle, introduce a time-dependent bias in the measured position of a spectral feature, particularly at late times as the SN fades and its contribution to the total observed flux decreases. This is most directly illustrated by the extraction of Image E (\autoref{sec:spectralextractions} and \autoref{sec:galaxies}), which was only recoverable at wavelengths $<$5500\,\AA, where the SN flux still exceeded that of the red lens galaxy G2.

A related systematic arises from the narrow absorption lines masked prior to the continuum fit. Strong host-galaxy interstellar lines are concentrated at observed wavelengths $<$5500\,\AA\, and predominantly affect Features 1–3, while telluric absorption bands and sky emission lines become more prevalent at longer wavelengths and affect Features 4–7. Joint fits using different combinations of features yield consistent time delays, suggesting that these effects introduce only minor biases in our measurements.

\subsection{From time-delays to $H_0$}
To connect our time-delay measurements with cosmology we use the lens model in \citet{Mortsell2026}, based on space-based imaging presented by \citet{Goobar2026}.

\citet{Mortsell2026} derive time-delays between the five lensed images, which scale inversely with $H_0$. Because the uncertainty on $H_0$ is determined primarily by the fractional precision of the time-delay measurements, the longest measured delay, $\Delta t_{AD}$, provides the strongest individual constraint. Using only $\Delta t_{AD}=-65.7 \pm 3.5$ days yields $H_0 = 69.3^{+8.1}_{-6.0}\;{\rm km\,s^{-1}\,Mpc^{-1}}$, for the free-slope lens model in \citet{Mortsell2026}. The larger fractional uncertainty on $H_0$ compared with $\Delta t_{AD}$ indicates that the current precision is not limited primarily by the time-delay measurement itself.

Combining all four spectroscopic time-delays yields $H_0 = 70.2^{+8.2}_{-6.1}\; {\rm km\,s^{-1}\,Mpc^{-1}}$, where the uncertainty is again dominated by the lens modeling; see \citet{Mortsell2026} for a discussion of the systematic uncertainties in the lens model.

The spectroscopic time-delays are consistent with the independent photometric time-delay analysis in \citet{Townsend2026}. 
They measure $\Delta t_{AB}^{\rm phot}=-10.6\pm2.4$ and $\Delta t_{AC}^{\rm phot}=1.2\pm2.6$ days, in excellent agreement with our results. Combining both spectroscopic and photometric time-delays gives $H_0 = 70.8^{+8.2}_{-6.1}\;{\rm km\,s^{-1}\,Mpc^{-1}}$ (for the free-slope lens model).

Because our method tracks the wavelength position of intrinsic spectral features rather than the broadband flux, it is largely insensitive to flux calibration uncertainties and does not require the multiple images to be simultaneously well sampled in common photometric passbands. It is also less sensitive to magnification perturbations introduced by microlensing, 
 which are likely affecting image A in the photometric time-delay analysis \citep{Townsend2026}. 

\section{Summary}\label{sec:summary}
We have presented a spectroscopic dataset and measurements of the temporal evolution of spectral features used to infer time-delays between the multiple images of the strongly lensed superluminous supernova SN\,2025wny. Unlike previous approaches that infer time-delays from broadband light curves or from individual spectral epochs, our method exploits the temporal evolution of multiple broad absorption and emission features measured from spatially resolved spectra obtained over several months.

We developed a framework to measure the positions of broad spectral features using Gaussian Process regression. The non-parametric reconstruction of each feature provides robust estimates of the line minima and maxima together with realistic uncertainties derived from the posterior predictive distribution, avoiding assumptions regarding the intrinsic line profile. These measurements form a homogeneous data set describing the spectroscopic evolution of each lensed image.

The time-delays are inferred through a joint fit to the temporal evolution of all measured spectral features. Each feature is modeled independently, while the relative delays between the lensed images are treated as global parameters shared by the entire data set. We explored both spline-based and Gaussian Process representations of the intrinsic feature evolution and sampled the posterior probability distribution of the delays using Markov Chain Monte Carlo techniques. The consistency between the independent models demonstrates that the inferred delays are not sensitive to the adopted description of the feature evolution.

Our analysis yields precise spectroscopic constraints on the relative arrival times of the multiple images (at $\sim$2 -- 3 days precision), with the longest delay $\Delta t_{AD} = -65.7 \pm 3.5$ days, measured to a precision of approximately 5\%. Because the uncertainty on $H_0$ is determined primarily by the fractional precision of the time-delay measurements, the longest measured delay ($\Delta t_{AD}$) provides the strongest individual constraint on $H_0 = 69.3^{+8.1}_{-6.0}\; {\rm km\,s^{-1}\,Mpc^{-1}}$, for the free-slope lens model in \citet{Mortsell2026}. Combining all four spectroscopic time-delays yields $H_0 = 70.2^{+8.2}_{-6.1}\; {\rm km\,s^{-1}\,Mpc^{-1}}$. 

The spectroscopic time-delays are consistent with the independent photometric time-delay analysis in \citet{Townsend2026}. Combining both spectroscopic and photometric time-delays gives $H_0 = 70.8^{+8.2}_{-6.1}\;{\rm km\,s^{-1}\,Mpc^{-1}}$ (for the free-slope lens model). It should be noted that a potential mass-sheet degeneracy can shift $H_0$ substantially. Future spatially resolved stellar velocity dispersion maps of the lens galaxies, together with a detailed characterization of the lens environment, will enable a robust determination of $H_0$.

In addition, future analyses could make use of spatially resolved photometry, to mangle and flux calibrate the spectra and make full use of the spectro-photometric information. This could also allow for better studies of potential systematics introduced by host- and lens galaxy light contamination.

The methodology presented here is entirely general and is applicable to future samples of strongly lensed supernovae expected from wide-field transient surveys such as LSST together with spectroscopic follow-up from large ground-based telescopes and the James Webb Space Telescope. As the sample of lensed supernovae grows, spectroscopic time-delay measurements will provide an important complement to photometric s, offering an independent route to measuring lensing delays, testing lens models, and ultimately improving cosmological constraints derived from time-delay cosmography.

\input{acknowledgments}

\facilities{Gemini:North (GMOS), GTC (OSIRIS+), HET (LRS2), Keck:I (LRIS), Keck:II (KCWI), MMT (Binospec), NOT (ALFOSC), PO:Hale (NGPS), UH88 (SNIFS), VLT:Antu (UT1 FORS2)}

\software{Astropy \citep{astropycollaboration_astropy:_2013, astropycollaboration_astropy_2018, AstropyCollaboration2022}, 
Matplotlib \citep{hunter_matplotlib:_2007}, 
}

\begin{table*}[htp]
    \centering
    \begin{tabular}{l l l l l l l l l}
        \hline
        \textbf{Image} & \textbf{MJD} & \textbf{Feature 1} & \textbf{Feature 2} & \textbf{Feature 3} & \textbf{Feature 4} & \textbf{Feature 5} & \textbf{Feature 6} & \textbf{Feature 7} \\
         & & $\lambda_{\rm max}$ (\AA) & $\lambda_{\rm min}$ (\AA) & $\lambda_{\rm min}$ (\AA) & $\lambda_{\rm min}$ (\AA) & $\lambda_{\rm min}$ (\AA) & $\lambda_{\rm min}$ (\AA) & $\lambda_{\rm min}$ (\AA) \\
        \hline
        \hline
A & 60944.164 & 4077.9 (17.8) & 4227.3 (40.7) & 5144.4 (32.4) & 5630.4 (34.0) & 6317.2 (56.0) & 7268.8 (44.0) & - \\
A & 60959.140 & 4060.3 (22.4) & 4265.1 (21.1) & 5205.0 (32.9) & 5653.9 (33.5) & 6397.8 (35.6) & 7077.3 (54.0) & 7705.9 (45.1) \\
A & 60963.548 & -- & 4246.8 (19.6) & -- & 5700.6 (45.6) & 6444.0 (24.2) & 7104.7 (32.4) & 7621.7 (21.6) \\
A & 60970.561 & 4092.8 (10.8) & 4241.1 (12.1) & 5239.1 (21.8) & 5620.7 (37.5) & 6476.4 (19.4) & 7110.8 (26.1) & 7780.3 (39.3) \\
A & 60979.371 & 4098.9 (10.5) & 4234.5 (19.1) & 5220.8 (26.5) & 5640.2 (40.3) & 6471.6 (22.2) & -- & -- \\
A & 60995.497 & -- & -- & 5271.7 (22.6) & 5722.0 (26.4) & 6379.8 (28.2) & 7153.0 (28.3) & 7955.3 (41.8) \\
A & 61004.520 & 4126.7 (6.9) & 4234.3( 10.2) & 5276.7 (29.1) & -- & -- & 7127.2 (24.9) & 7924.4 (38.5) \\
A & 61008.073 & 4135.6 (5.8) & 4250.0 (13.3) & 5239.6 (28.4) & 5713.8 (19.5) & 6373.3 (26.1) & 7155.4 (25.7) & 7896.0 (39.4) \\
A & 61008.346 & 4134.9 (5.8) & 4234.4 (12.9) & 5277.0 (28.0) & 5714.5 (20.4) & 6409.3 (31.7) & 7110.4 (20.3) & 7776.0 (97.2) \\
A & 61024.270 & 4156.1 (6.7) & 4269.8 (12.4) & 5323.9 (48.4) & 5729.5 (23.1) & 6389.3 (25.8) & 7168.6 (25.3) & 8123.2 (78.4) \\
A & 61027.400 & 4160.9 (4.9) & 4266.3 (9.0) & 5318.4 (23.5) & 5716.9 (20.3) & 6398.9 (22.7) & 7137.4 (26.6) & 8011.2 (38.9) \\
A & 61030.339 & 4164.1 (6.3) & 4277.3 (10.6) & 5331.6 (28.1) & 5742.9 (19.8) & 6415.9 (23.6) & 7136.6 (25.0) & 8003.2 (40.3) \\
A & 61033.262 & 4167.2 (7.8) & 4284.1 (9.8) & 5321.4 (23.6) & 5765.4 (31.9) & 6410.9 (21.9) & 7199.4 (25.5) & 8102.1 (83.5) \\
A & 61051.000 & 4183.9 (5.4) & 4309.9 (8.8) & -- & 5850.7 (20.5) & 6444.0 (18.6) & 7138.2 (19.4) & 8066.8 (23.4) \\
A & 61058.160 & 4186.8 (8.0) & 4328.8 (13.6) & 5376.3 (36.2) & 5794.7 (36.0) & 6471.4 (20.1) & 7232.4 (22.4) & 8180.9 (36.4) \\
A & 61060.360 & 4188.5 (4.7) & 4333.7 (8.8) & 5369.7 (44.1) & 5834.4 (23.5) & 6455.2 (18.7) & 7217.2 (24.8) & -- \\
A & 61080.080 & 4209.1 (9.5) & 4344.5 (13.8) & 5377.1 (27.7) & 5841.5 (23.2) & 6481.9 (17.4) & 7179.6 (21.9) & 8167.6 (57.7) \\
A & 61090.258 & 4214.5 (5.4) & 4361.2 (12.4) & 5407.1 (26.2) & 5853.8 (23.4) & 6484.5 (20.7) & 7245.8 (78.9) & 8075.4 (37.6) \\
A & 61114.010 & 4220.9 (7.8) & 4387.0 (12.0) & 5395.3 (37.4) & 5873.3 (19.4) & 6498.8 (18.1) & -- & -- \\
A & 61138.020 & 4238.2 (11.7) & 4391.0 (14.9) & 5393.3 (36.2) & 5878.4 (30.3) & -- & -- & -- \\
B & 60970.561 & 4104.2 (9.5) & 4216.9 (14.8) & 5233.5 (52.2) & 5683.9 (31.9) & 6462.2 (46.0) & 7085.0 (22.5) & 7841.1 (63.9) \\
B & 60979.371 & 4110.2 (17.1) & 4267.2 (23.2) & 5263.2 (52.0) & 5764.5 (23.3) & 6325.2 (55.5) & -- & -- \\
B & 60995.497 & -- & -- & -- & 5722.1 (28.6) & 6344.8 (37.3) & 7136.2 (39.7) & 7992.1 (55.9) \\
B & 61004.520 & 4140.7 (6.8) & 4239.0 (12.5) & 5315.7 (42.4) & -- & -- & 7126.1 (28.4) & 7993.3 (45.8) \\
B & 61024.270 & 4165.6 (8.5) & 4284.4 (17.3) & 5360.1 (66.7) & 5772.6 (48.9) & 6482.8 (101.2) & 7151.2 (36.5) & 8176.8 (102.0) \\
B & 61027.400 & 4173.1 (8.9) & 4295.3 (13.5) & 5348.2 (32.1) & 5764.8 (48.7) & 6436.6 (37.3) & 7135.6 (54.9) & 8023.1 (66.5) \\
B & 61030.339 & 4176.7 (10.5) & 4309.9 (15.9) & 5363.5 (28.3) & 5754.4 (57.4) & 6447.2 (40.8) & 7118.4 (50.1) & 8056.2 (95.7) \\
B & 61051.000 & 4192.6 (6.6) & 4338.2 (17.4) & -- & 5848.7 (26.9) & 6464.0 (21.4) & 7127.6 (23.4) & 8087.9 (52.1) \\
B & 61058.160 & 4195.4 (9.6) & 4320.4 (15.8) & 5421.2 (32.9) & 5879.8 (32.3) & 6461.9 (34.3) & 7203.0 (31.9) & 8188.4 (129.0) \\
B & 61060.360 & 4197.1 (7.3) & 4325.4 (17.0) & 5211.4 (21.1) & 5778.0 (16.5) & 6423.9 (31.6) & 7269.8 (26.2) & -- \\
B & 61080.080 & 4207.6 (8.2) & 4380.0 (27.7) & -- & -- & -- & -- & -- \\
B & 61090.258 & 4211.8 (8.0) & 4356.6 (14.8) & 5388.2 (33.9) & 5837.7 (26.6) & 6497.8(44.2) & -- & -- \\
B & 61114.010 & 4227.4 (9.5) & 4402.0 (18.7) & 5376.9 (37.0) & 5917.4 (50.3) & -- & -- & -- \\
B & 61138.020 & 4221.2 (13.0) & 4399.5 (15.2) & 5339.7 (37.5) & -- & -- & -- & -- \\
C & 60970.561 & 4086.6 (12.9) & 4243.1 (16.0) & 5239.4 (25.6) & -- & 6441.4(35.6) & 7100.1 (24.1) & 7771.9 (87.8) \\
C & 60979.371 & 4081.8 (33.0) & 4223.3 (33.9) & 5137.6 (79.4) & 5863.4 (59.2) & 6312.0(78.0) & -- & -- \\
C & 61004.000 & -- & 4243.6 (17.2) & 5275.0 (60.0) & -- & -- & 7088.0 (23.5) & 7945.7 (52.4) \\
C & 61004.520 & 4131.6 (7.0) & -- & -- & -- & - & -- & -- \\
C & 61051.000 & 4182.1 (6.9) & 4307.2 (16.0) & -- & 5800.7 (88.6) & 6457.6 (23.0) & 7109.8 (23.4) & 8092.0 (33.2) \\
D & 60970.561 & 4167.7 (8.7) & 4290.8 (14.8) & 5372.5 (58.4) & 5775.3 (41.9) & 6390.8 (38.1) & 7091.9 (57.5) & 8058.4 (154.2) \\
D & 61004.520 & 4197.2 (6.9) & 4343.9 (14.8) & 5435.2 (30.0) & -- & -- & -- & 8122.2 (100.0) \\
D & 61008.073 & 4197.4 (8.1) & 4339.5 (16.6) & 5346.7 (39.4) & 5856.3 (50.0) & 6464.7 (58.9) & 7182.0 (65.7) & 8164.9 (126.3) \\
D & 61051.000 & 4221.9 (10.8) & 4388.4 (14.8) & -- & 5849.3 (89.4) & 6491.2 (89.9) & -- & 8143.3 (60.0) \\
D & 61060.360 & 4233.7 (11.8) & 4401.7 (20.0) & 5442.3 (19.3) & 5912.0 (16.7) & 6539.9 (20.8) & 7178.0 (28.7) & -- \\
E & 60970.561 & 4048.7 (20.7) & 4197.9 (33.8) & -- & -- & -- & -- & -- \\
E & 61004.520 & 4134.9 (13.7) & 4247.3 (17.8) & 5298.0 (53.2) & -- & -- & -- & -- \\
\hline
\end{tabular}
    \caption{Measured wavelengths (observer-frame) of the broad emission/absorption features 1--7 for images A--E. Missing measurements are indicated by "--".}
    \label{tab:features}
\end{table*}

\clearpage
\appendix
\section{Redshift and velocity dispersion measurements of the lens galaxies}\label{sec:galaxies}
We measured the stellar velocity dispersions of the two foreground lens galaxies by fitting the available spectra with the stellar template fitting code \texttt{pPXF} \citep{Cappellari2023}, which implements penalised pixel-fitting to extract the moments of the line-of-sight velocity distribution from galaxy spectra. This method works by searching a library of template spectra over a range of metallicities and ages, in this case an ensemble of simple stellar populations (SSPs), to fit to the observed spectra and thus deliver robust kinematic measurements. The template library is taken from the E-MILES stellar population models \citep{Vazdekis2016MNRAS.463.3409V}, chosen for their broad spectral range, good resolution (FWHM = 2.5~\AA\,from 3540 to 8950~\AA) and extensive coverage in age and metallicity ($-1.79 < [\mathrm{M/H}] < +0.26$ and ages $>30$ Myr).

To account for possible contamination from the background host galaxy or cross-contamination between G1 and G2 and the bright SN images, we employ a multi-component fitting technique following \citet{turner2024MNRAS.528.3559T}. In this framework, the observed spectrum is modeled as the sum of distinct spectral components, each assigned its own redshift and kinematics, so that contaminating continuum or emission features are not absorbed into the stellar component of the foreground lens galaxies. 

As in the multiple-component fits of \citet{turner2024MNRAS.528.3559T}, the purpose of the additional components is not primarily to interpret their own kinematics, but to prevent contaminating spectral features from biasing the recovered foreground-galaxy velocity dispersion. We therefore selected the fiducial models by requiring both a good total fit and a stable, physically sensible lens-galaxy component, rather than simply adopting the most complex model with the lowest formal $\chi^2$.

From our highest resolution Keck/KCWI spectra ($R\sim3600$ and $R\sim2000$ for the blue and red arm, respectively) we  measure the redshifts for the two lens galaxies $z_{G1} = 0.37561 \pm 0.00002$ and $z_{G2} = 0.37669 \pm 0.00003$. The corresponding rest-frame velocity difference is $\Delta v=236\pm9~\mathrm{km\,s^{-1}}$, with G2 redshifted relative to G1.

The best-fit stellar velocity dispersion for G1 is $\sigma_{v}^{\rm G1, KCWI} = 289.1\pm4.8~\mathrm{km\,s^{-1}}$, which is consistent with the result from the DESI spectrum, $\sigma_{v}^{\rm G1, DESI} = 298\pm37~\mathrm{km\,s^{-1}}$ in \citet{storfer2026arXiv260402418S}. For G2, we measure $\sigma_{v}^{\rm G2, KCWI} = 184.2\pm8.8~\mathrm{km\,s^{-1}}$. These velocity dispersions provide independent spectroscopic constraints on the mass distribution of the lens galaxies and are subsequently used in the lens modeling by \citet{Mortsell2026}.
\begin{figure*}[htp]
    \centering
    \includegraphics[width=\textwidth]{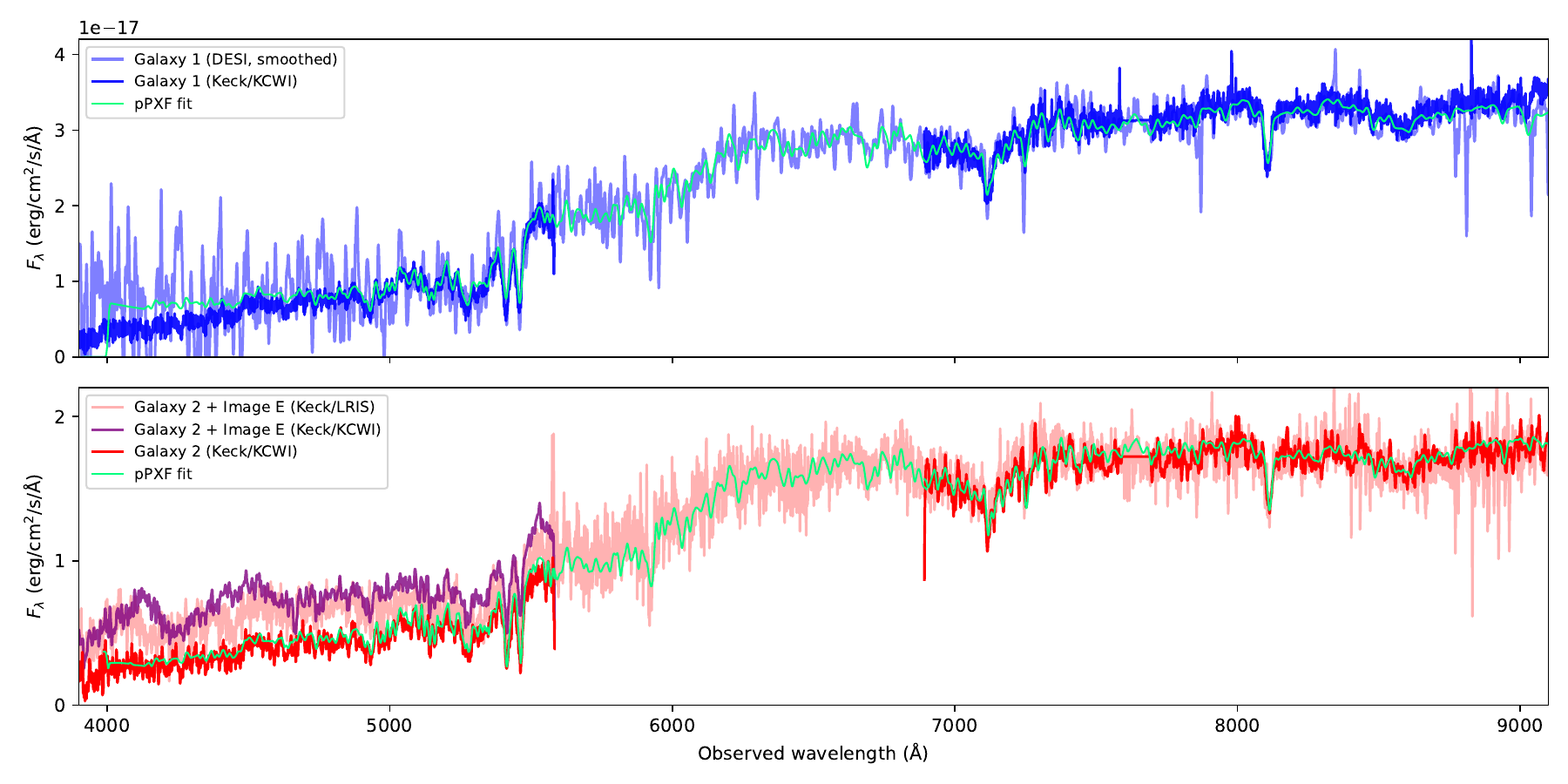}
    \caption{
    Spectra of the lens galaxies G1 (top panel) and G2 (bottom panel). For G1 we use both the DESI (light blue line) and Keck/KCWI (dark blue) spectra to fit a stellar velocity dispersion and redshift. Similarly, for G2 we use long-slit spectra from Keck/LRIS (pink lines, which is blended with supernova flux from image E) and blended/de-blended spectra from Keck/KCWI (purple/red lines, respectively). The pPXF best-fit models are shown as green lines.
    }
    \label{fig:image_e}
\end{figure*}

\normalsize
\clearpage

\bibliography{references, Lensbib}

\end{document}

%% file: affiliations.tex
\newcommand{\WIS}{\affiliation{Department of Particle Physics and Astrophysics, Weizmann Institute of Science, 76100 Rehovot, Israel}}

\newcommand{\CSIC}{\affiliation{Instituto de Astrof\'{\i}sica de Andaluc\'{\i}a CSIC, Granada, E-18008, Spain}}

\newcommand{\ICECSIC}{\affiliation{Institute of Space Sciences (ICE-CSIC), Campus UAB, Carrer de Can Magrans, s/n, E-08193 Barcelona, Spain.}}

\newcommand{\IEEC}{\affiliation{Institut d'Estudis Espacials de Catalunya (IEEC), 08860 Castelldefels (Barcelona), Spain}}

\newcommand{\LJMU}{\affiliation{Astrophysics Research Institute, Liverpool John Moores University, 146 Brownlow Hill, Liverpool L3 5RF, UK}}

\newcommand{\birmingham}{\affiliation{School of Physics \& Astronomy and Institute of Gravitational Wave Astronomy, University of Birmingham, Edgbaston, B15 2TT, UK}}

\newcommand{\HUBerlin}{\affiliation{Institut f\"ur Physik, Humboldt-Universit\"at zu Berlin, Newtonstr. 15, 12489 Berlin, Germany}}

\newcommand{\OKC}{\affiliation{Department of Physics, Oskar Klein Centre, Stockholm University, SE-106 91, Stockholm, Sweden}}
\newcommand{\OKCAstro}{\affiliation{Department of Astronomy, Oskar Klein Center, Stockholm University, SE-106 91 Stockholm, Sweden}}
\newcommand{\Caltech}{\affiliation{Cahill Center for Astronomy and Astrophysics, California Institute of Technology, Mail Code 249-17, Pasadena, CA 91125, USA}}
\newcommand{\CaltechPhys}{\affiliation{Division of Physics, Mathematics and Astronomy, California Institute of Technology, Pasadena, CA 91125, USA}}

\newcommand{\CIERA}{\affiliation{Center for Interdisciplinary Exploration and Research in Astrophysics (CIERA), 1800 Sherman Ave., Evanston, IL 60201, USA}}
\newcommand{\Northwestern}{\affiliation{Department of Physics and Astronomy, Northwestern University, 2145 Sheridan Rd, Evanston, IL 60208, USA}}
\newcommand{\SkAI}{\affiliation{NSF-Simons AI Institute for the Sky (SkAI), 172 E. Chestnut St., Chicago, IL 60611, USA}}

\newcommand{\CaltechOO}{\affiliation{Caltech Optical Observatories, California Institute of Technology, Pasadena, CA 91125, USA}}

\newcommand{\LMU}{\affiliation{University Observatory, Faculty of Physics, Ludwig-Maximilians-Universität, Scheinerstr. 1, 81679 Munich, Germany}}
\newcommand{\ORIGINS}{\affiliation{Excellence Cluster ORIGINS, Boltzmannstr. 2, 85748 Garching, Germany}}
\newcommand{\UCLA}{\affiliation{Department of Physics and Astronomy, UCLA PAB 430 Portola Plaza Los Angeles, CA 90095-1547}}

%% file: authors_25wny_td.tex
\author[0000-0001-5975-290X]{Joel~Johansson}
\OKC

\author[0000-0002-8380-6143]{Edvard~Mörtsell}
\OKC

\author[0000-0002-4163-4996]{Ariel~Goobar}
\OKC

\author[0000-0001-6797-1889]{Steve~Schulze}
\WIS

\author[0009-0001-6911-9144]{Maggie~L.~Li}
\Caltech

\author[0000-0003-3658-6026]{Yu-Jing~Qin}
\Caltech

\author[0000-0003-1710-9339]{Lin~Yan}
\CaltechPhys
\CaltechOO

\author[0009-0006-7102-3674]{Hannah~C.~Turner}
\birmingham

\author[0000-0002-2376-6979]{Suhail~Dhawan}
\birmingham

\author[0000-0001-6343-3362]{Alice~Townsend}
\birmingham

\author[0000-0001-8342-6274]{Jakob~Nordin}
\HUBerlin


\author[0009-0008-2714-2507]{Aleksandra~Bochenek}
\LJMU

\author[0009-0001-0574-2332]{Malte~Busmann}
\LMU
\ORIGINS

\author[0000-0001-8372-997X]{Kaustav~K.~Das}
\Caltech

\author[0000-0002-4223-103X]{Christoffer~Fremling}
\CaltechPhys
\CaltechOO

\author[0000-0002-1296-6887]{LLuis~Galbany}
\ICECSIC
\IEEC

\author[0000-0002-5025-4645]{Alexa~C.~Gordon}
\CIERA

\author[0000-0003-3270-7644]{Daniel~Gruen}
\LMU
\ORIGINS

\author[0000-0002-5619-4938]{Mansi~M.~Kasliwal}
\Caltech

\author[0000-0002-7866-4531]{Chang~Liu}
\Northwestern
\CIERA
\SkAI

\author[0009-0006-0726-1328]{Zoë~McGrath}
\LJMU

\author[0000-0002-8650-1644]{Christopher~Martin}
\Caltech

\author[0009-0005-8228-0329]{Peter~Massey}
\birmingham

\author[0000-0002-0610-2644]{Martijn~S.~S.~L.~Oei}
\Caltech

\author[0000-0001-8472-1996]{Daniel~A.~Perley}
\LJMU

\author[0009-0005-7030-8742]{Andr\'es~I.~Ponte~P\'erez}
\birmingham

\author[0000-0001-7145-8674]{Francisco~Prada}
\CSIC

\author[0000-0001-5847-7934]{Nikolaus~Z.~Prusinski}
\Caltech

\author[0000-0002-5683-2389]{Nabeel~Rehemtulla}
\Northwestern
\CIERA
\SkAI

\author[0000-0003-0427-8387]{R.~Michael~Rich}
\UCLA

\author[0009-0006-1510-4648]{Surya~Shivaprasad}
\LMU
\ORIGINS

\author[0000-0003-1546-6615]{Jesper~Sollerman}
\OKCAstro

\author[0000-0003-0733-2916]{Jacob~L.~Wise}
\LJMU

%% file: acknowledgments.tex
\section*{Acknowledgements}
We thank Aswin Suresh, Yuxin Dong, Christoph Ries, Michael Schmidt, and Silona Wilke for carrying out observations. 

E.M.\ acknowledges support from the Swedish
Research Council under Dnr VR 2024-03927.

A.G.\ acknowledges support from the Swedish Research Council through project Dnr 2020-03444 and the Swedish National Space Agency, Dnr 2023-00226.

S.D., H.C.T., and A.T.\ acknowledge support from  UK Research and Innovation (UKRI) under the UK government’s Horizon Europe funding Guarantee EP/Z000475/1.

A.C.G. and the Fong Group at Northwestern acknowledge support by the National Science Foundation under grant Nos. AST-1909358, AST-2206494, AST-2308182 and CAREER grant No. AST-2047919. 

C.L. is supported by DoE award \#\,DE-SC0025599.
W. M. Keck Observatory and MMT Observatory access was supported by Northwestern University and the Center for Interdisciplinary Exploration and Research in Astrophysics (CIERA). 

Funded in part by the Deutsche Forschungsgemeinschaft (DFG, German Research Foundation) under Germany's Excellence Strategy – EXC-2094/2 – 390783311.

Based on observations obtained with the Samuel Oschin Telescope 48-inch and the 60-inch Telescope at the Palomar Observatory as part of the Zwicky Transient Facility project. ZTF is supported by the National Science Foundation under Award 2407588 and a partnership including Caltech, USA; Caltech/IPAC, USA; University of Maryland, USA; University of California, Berkeley, USA; University of Wisconsin at Milwaukee, USA; Cornell University, USA; Drexel University, USA; University of North Carolina at Chapel Hill, USA; Institute of Science and Technology, Austria; National Central University, Taiwan, and OKC, University of Stockholm, Sweden. Operations are conducted by Caltech's Optical Observatory (COO), Caltech/IPAC, and the University of Washington at Seattle, USA. 

Based on observations made with the Nordic Optical Telescope, owned in collaboration by the University of Turku and Aarhus University, and operated jointly by Aarhus University, the University of Turku and the University of Oslo, representing Denmark, Finland and Norway, the University of Iceland and Stockholm University at the Observatorio del Roque de los Muchachos, La Palma, Spain, of the Instituto de Astrofisica de Canarias. The NOT data were obtained under program ID P70-501.

Based on observations made with the Gran Telescopio Canarias (GTC), installed at the Spanish Observatorio del Roque de los Muchachos of the Instituto de Astrofísica de Canarias, on the island of La Palma. This work partially based on data obtained with the instrument OSIRIS$+$, built by a Consortium led by the Instituto de Astrofísica de Canarias in collaboration with the Instituto de Astronomía of the Universidad Autónoma de México. OSIRIS was funded by GRANTECAN and the National Plan of Astronomy and Astrophysics of the Spanish Government.

Some of the data presented herein were obtained at the W.~M. Keck Observatory, which is operated as a scientific partnership among the California Institute of Technology, the University of California, and NASA. The Observatory was made possible by the generous financial support of the W.~M. Keck Foundation. The authors wish to recognize and acknowledge the very significant cultural role and reverence that the summit of Maunakea has always had within the indigenous Hawaiian community. We are most fortunate to have the opportunity to conduct observations from this mountain.

This paper contains data from observations obtained with the Hobby-Eberly Telescope (HET), which is a joint project of the University of Texas at Austin, the Pennsylvania State University, Ludwig-Maximilians-Universität München, and Georg-August Universität Göttingen. The HET is named in honor of its principal benefactors, William P. Hobby and Robert E. Eberly. We acknowledge the Texas Advanced Computing Center (TACC) at The University of Texas at Austin for providing high-performance computing, visualization, and storage resources that have contributed to the results reported within this paper. The Low Resolution Spectrograph 2 (LRS2) was developed and funded by the University of Texas at Austin, McDonald Observatory, Department of Astronomy, and Pennsylvania State University. We thank the Leibniz-Institut für Astrophysik Potsdam (AIP) and the Institut für Astrophysik Göttingen (IAG) for their contributions to the construction of the integral field units.

%% file: main.bbl
\begin{thebibliography}{}
\expandafter\ifx\csname natexlab\endcsname\relax\def\natexlab#1{#1}\fi
\providecommand{\url}[1]{\href{#1}{#1}}
\providecommand{\dodoi}[1]{doi:~\href{http://doi.org/#1}{\nolinkurl{#1}}}
\providecommand{\doeprint}[1]{\href{http://ascl.net/#1}{\nolinkurl{http://ascl.net/#1}}}
\providecommand{\doarXiv}[1]{\href{https://arxiv.org/abs/#1}{\nolinkurl{https://arxiv.org/abs/#1}}}

\bibitem[{{Appenzeller} {et~al.}(1998){Appenzeller}, {Fricke}, {F{\"u}rtig}, {G{\"a}ssler}, {H{\"a}fner}, {Harke}, {Hess}, {Hummel}, {J{\"u}rgens}, {Kudritzki}, {Mantel}, {Meisl}, {Muschielok}, {Nicklas}, {Rupprecht}, {Seifert}, {Stahl}, {Szeifert}, \& {Tarantik}}]{Appenzeller1998Msngr..94....1A}
{Appenzeller}, I., {Fricke}, K., {F{\"u}rtig}, W., {et~al.} 1998, The Messenger, 94, 1

\bibitem[{{Astropy Collaboration} {et~al.}(2013){Astropy Collaboration}, Robitaille, Tollerud, Greenfield, Droettboom, Bray, Aldcroft, Davis, Ginsburg, {Price-Whelan}, Kerzendorf, Conley, Crighton, Barbary, Muna, Ferguson, Grollier, Parikh, Nair, G{\"u}nther, Deil, Woillez, Conseil, Kramer, Turner, Singer, Fox, Weaver, Zabalza, Edwards, Azalee~Bostroem, Burke, Casey, Crawford, Dencheva, Ely, Jenness, Labrie, Lim, Pierfederici, Pontzen, Ptak, Refsdal, Servillat, \& Streicher}]{astropycollaboration_astropy:_2013}
{Astropy Collaboration}, Robitaille, T.~P., Tollerud, E.~J., {et~al.} 2013, A\&A, 558, A33, \dodoi{10.1051/0004-6361/201322068}

\bibitem[{{Astropy Collaboration} {et~al.}(2018){Astropy Collaboration}, {Price-Whelan}, Sip{\H o}cz, G{\"u}nther, Lim, Crawford, Conseil, Shupe, Craig, Dencheva, Ginsburg, VanderPlas, Bradley, {P{\'e}rez-Su{\'a}rez}, {de Val-Borro}, Aldcroft, Cruz, Robitaille, Tollerud, Ardelean, Babej, Bachetti, Bakanov, Bamford, Barentsen, Barmby, Baumbach, Berry, Biscani, Boquien, Bostroem, Bouma, Brammer, Bray, Breytenbach, Buddelmeijer, Burke, Calderone, Rodr{\'i}guez, Cara, Cardoso, Cheedella, Copin, Crichton, D'{\'A}vella, Deil, Depagne, Dietrich, Donath, Droettboom, Earl, Erben, Fabbro, Ferreira, Finethy, Fox, Garrison, Gibbons, Goldstein, Gommers, Greco, Greenfield, Groener, Grollier, Hagen, Hirst, Homeier, Horton, Hosseinzadeh, Hu, Hunkeler, Ivezi{\'c}, Jain, Jenness, Kanarek, Kendrew, Kern, Kerzendorf, Khvalko, King, Kirkby, Kulkarni, Kumar, Lee, Lenz, Littlefair, Ma, Macleod, Mastropietro, McCully, Montagnac, Morris, Mueller, Mumford, Muna, Murphy, Nelson, Nguyen, Ninan, N{\"o}the, Ogaz, Oh, Parejko, Parley,
  Pascual, Patil, Patil, Plunkett, Prochaska, Rastogi, Janga, Sabater, Sakurikar, Seifert, Sherbert, {Sherwood-Taylor}, Shih, Sick, Silbiger, Singanamalla, Singer, Sladen, Sooley, Sornarajah, Streicher, Teuben, Thomas, Tremblay, Turner, Terr{\'o}n, {van Kerkwijk}, {de la Vega}, Watkins, Weaver, Whitmore, Woillez, \& Zabalza}]{astropycollaboration_astropy_2018}
{Astropy Collaboration}, {Price-Whelan}, A.~M., Sip{\H o}cz, B.~M., {et~al.} 2018, AJ, 156, 123, \dodoi{10.3847/1538-3881/aabc4f}

\bibitem[{{Astropy Collaboration} {et~al.}(2022){Astropy Collaboration}, {Price-Whelan}, {Lim}, {Earl}, {Starkman}, {Bradley}, {Shupe}, {Patil}, {Corrales}, {Brasseur}, {N{\"o}the}, {Donath}, {Tollerud}, {Morris}, {Ginsburg}, {Vaher}, {Weaver}, {Tocknell}, {Jamieson}, {van Kerkwijk}, {Robitaille}, {Merry}, {Bachetti}, {G{\"u}nther}, {Aldcroft}, {Alvarado-Montes}, {Archibald}, {B{\'o}di}, {Bapat}, {Barentsen}, {Baz{\'a}n}, {Biswas}, {Boquien}, {Burke}, {Cara}, {Cara}, {Conroy}, {Conseil}, {Craig}, {Cross}, {Cruz}, {D'Eugenio}, {Dencheva}, {Devillepoix}, {Dietrich}, {Eigenbrot}, {Erben}, {Ferreira}, {Foreman-Mackey}, {Fox}, {Freij}, {Garg}, {Geda}, {Glattly}, {Gondhalekar}, {Gordon}, {Grant}, {Greenfield}, {Groener}, {Guest}, {Gurovich}, {Handberg}, {Hart}, {Hatfield-Dodds}, {Homeier}, {Hosseinzadeh}, {Jenness}, {Jones}, {Joseph}, {Kalmbach}, {Karamehmetoglu}, {Ka{\l}uszy{\'n}ski}, {Kelley}, {Kern}, {Kerzendorf}, {Koch}, {Kulumani}, {Lee}, {Ly}, {Ma}, {MacBride}, {Maljaars}, {Muna}, {Murphy}, {Norman},
  {O'Steen}, {Oman}, {Pacifici}, {Pascual}, {Pascual-Granado}, {Patil}, {Perren}, {Pickering}, {Rastogi}, {Roulston}, {Ryan}, {Rykoff}, {Sabater}, {Sakurikar}, {Salgado}, {Sanghi}, {Saunders}, {Savchenko}, {Schwardt}, {Seifert-Eckert}, {Shih}, {Jain}, {Shukla}, {Sick}, {Simpson}, {Singanamalla}, {Singer}, {Singhal}, {Sinha}, {Sip{\H{o}}cz}, {Spitler}, {Stansby}, {Streicher}, {{\v{S}}umak}, {Swinbank}, {Taranu}, {Tewary}, {Tremblay}, {de Val-Borro}, {Van Kooten}, {Vasovi{\'c}}, {Verma}, {de Miranda Cardoso}, {Williams}, {Wilson}, {Winkel}, {Wood-Vasey}, {Xue}, {Yoachim}, {Zhang}, {Zonca}, \& {Astropy Project Contributors}}]{AstropyCollaboration2022}
{Astropy Collaboration}, {Price-Whelan}, A.~M., {Lim}, P.~L., {et~al.} 2022, \apj, 935, 167, \dodoi{10.3847/1538-4357/ac7c74}

\bibitem[{{Bayer} {et~al.}(2021){Bayer}, {Huber}, {Vogl}, {Suyu}, {Taubenberger}, {Sluse}, {Chan}, \& {Kerzendorf}}]{Bayer2021A&A...653A..29B}
{Bayer}, J., {Huber}, S., {Vogl}, C., {et~al.} 2021, \aap, 653, A29, \dodoi{10.1051/0004-6361/202040169}

\bibitem[{{Cappellari}(2023)}]{Cappellari2023}
{Cappellari}, M. 2023, MNRAS, 526, 3273, \dodoi{10.1093/mnras/stad2597}

\bibitem[{{Cepa} {et~al.}(2000){Cepa}, {Aguiar}, {Escalera}, {Gonzalez-Serrano}, {Joven-Alvarez}, {Peraza}, {Rasilla}, {Rodriguez-Ramos}, {Gonzalez}, {Cobos Duenas}, {Sanchez}, {Tejada}, {Bland-Hawthorn}, {Militello}, \& {Rosa}}]{Cepa2000SPIE.4008..623C}
{Cepa}, J., {Aguiar}, M., {Escalera}, V.~G., {et~al.} 2000, in Society of Photo-Optical Instrumentation Engineers (SPIE) Conference Series, Vol. 4008, Optical and IR Telescope Instrumentation and Detectors, ed. M.~{Iye} \& A.~F. {Moorwood}, 623--631, \dodoi{10.1117/12.395520}

\bibitem[{Chen {et~al.}(2022)Chen, Kelly, Oguri, Broadhurst, Diego, Emami, Filippenko, Treu, \& Zitrin}]{chen_2022}
Chen, W., Kelly, P.~L., Oguri, M., {et~al.} 2022, Nature, 611, 256, \dodoi{10.1038/s41586-022-05252-5}

\bibitem[{{Chen} {et~al.}(2024){Chen}, {Kelly}, {Frye}, {Pierel}, {Willner}, {Pascale}, {Cohen}, {Conselice}, {Engesser}, {Furtak}, {Gilman}, {Grogin}, {Huber}, {Jha}, {Johansson}, {Koekemoer}, {Larison}, {Meena}, {Siebert}, {Windhorst}, {Yan}, \& {Zitrin}}]{Chen2024ApJ...970..102C}
{Chen}, W., {Kelly}, P.~L., {Frye}, B.~L., {et~al.} 2024, \apj, 970, 102, \dodoi{10.3847/1538-4357/ad50a5}

\bibitem[{{Chonis} {et~al.}(2014){Chonis}, {Hill}, {Lee}, {Tuttle}, \& {Vattiat}}]{Chonis2014}
{Chonis}, T.~S., {Hill}, G.~J., {Lee}, H., {Tuttle}, S.~E., \& {Vattiat}, B.~L. 2014, in Society of Photo-Optical Instrumentation Engineers (SPIE) Conference Series, Vol. 9147, Ground-based and Airborne Instrumentation for Astronomy V, ed. S.~K. {Ramsay}, I.~S. {McLean}, \& H.~{Takami}, 91470A, \dodoi{10.1117/12.2056005}

\bibitem[{{Chonis} {et~al.}(2016){Chonis}, {Hill}, {Lee}, {Tuttle}, {Vattiat}, {Drory}, {Indahl}, {Peterson}, \& {Ramsey}}]{Chonis2016}
{Chonis}, T.~S., {Hill}, G.~J., {Lee}, H., {et~al.} 2016, in Society of Photo-Optical Instrumentation Engineers (SPIE) Conference Series, Vol. 9908, Ground-based and Airborne Instrumentation for Astronomy VI, ed. C.~J. {Evans}, L.~{Simard}, \& H.~{Takami}, 99084C, \dodoi{10.1117/12.2232209}

\bibitem[{{Dhawan} {et~al.}(2020){Dhawan}, {Johansson}, {Goobar}, {Amanullah}, {M{\"o}rtsell}, {Cenko}, {Cooray}, {Fox}, {Goldstein}, {Kalender}, {Kasliwal}, {Kulkarni}, {Lee}, {Nayyeri}, {Nugent}, {Ofek}, \& {Quimby}}]{Dhawan2020MNRAS.491.2639D}
{Dhawan}, S., {Johansson}, J., {Goobar}, A., {et~al.} 2020, \mnras, 491, 2639, \dodoi{10.1093/mnras/stz2965}

\bibitem[{{Fabricant} {et~al.}(2019){Fabricant}, {Fata}, {Epps}, {Gauron}, {Mueller}, {Zajac}, {Amato}, {Barberis}, {Bergner}, {Brennan}, {Brown}, {Chilingarian}, {Geary}, {Kradinov}, {McLeod}, {Smith}, \& {Woods}}]{Fabricant2019PASP..131g5004F}
{Fabricant}, D., {Fata}, R., {Epps}, H., {et~al.} 2019, \pasp, 131, 075004, \dodoi{10.1088/1538-3873/ab1d78}

\bibitem[{{Foreman-Mackey} {et~al.}(2013){Foreman-Mackey}, Hogg, Lang, \& Goodman}]{foreman-mackey_emcee_2013}
{Foreman-Mackey}, D., Hogg, D.~W., Lang, D., \& Goodman, J. 2013, PASP, 125, 306, \dodoi{10.1086/670067}

\bibitem[{{Frye} {et~al.}(2024){Frye}, {Pascale}, {Pierel}, {Chen}, {Foo}, {Leimbach}, {Garuda}, {Cohen}, {Kamieneski}, {Windhorst}, {Koekemoer}, {Kelly}, {Summers}, {Engesser}, {Liu}, {Furtak}, {Polletta}, {Harrington}, {Willner}, {Diego}, {Jansen}, {Coe}, {Conselice}, {Dai}, {Dole}, {D'Silva}, {Driver}, {Grogin}, {Marshall}, {Meena}, {Nonino}, {Ortiz}, {Pirzkal}, {Robotham}, {Ryan}, {Strolger}, {Tompkins}, {Willmer}, {Yan}, {Yun}, \& {Zitrin}}]{Frye2024ApJ...961..171F}
{Frye}, B.~L., {Pascale}, M., {Pierel}, J., {et~al.} 2024, \apj, 961, 171, \dodoi{10.3847/1538-4357/ad1034}

\bibitem[{{Goobar} {et~al.}(2026){Goobar}, {Johansson}, {M\"ortsell}, \& {others}}]{Goobar2026}
{Goobar}, A., {Johansson}, J., {M\"ortsell}, E., \& {others}. 2026, ApJ, submitted

\bibitem[{{Goobar} {et~al.}(2025){Goobar}, {Johansson}, \& {Sagu{\'e}s Carracedo}}]{Goobar2025RSPTA.38340123G}
{Goobar}, A., {Johansson}, J., \& {Sagu{\'e}s Carracedo}, A. 2025, Philosophical Transactions of the Royal Society of London Series A, 383, 20240123, \dodoi{10.1098/rsta.2024.0123}

\bibitem[{Goobar {et~al.}(2017)Goobar, Amanullah, Kulkarni, Nugent, Johansson, Steidel, Law, Mörtsell, Quimby, Blagorodnova, Brandeker, Cao, Cooray, Ferretti, Fremling, Hangard, Kasliwal, Kupfer, Lunnan, Masci, Miller, Nayyeri, Neill, Ofek, Papadogiannakis, Petrushevska, Ravi, Sollerman, Sullivan, Taddia, Walters, Wilson, Yan, \& Yaron}]{goobar_2017}
Goobar, A., Amanullah, R., Kulkarni, S.~R., {et~al.} 2017, Science, 356, 291, \dodoi{10.1126/science.aal2729}

\bibitem[{Goobar {et~al.}(2023)Goobar, Johansson, Schulze, Arendse, Carracedo, Dhawan, Mörtsell, Fremling, Yan, Perley, Sollerman, Joseph, Hinds, Meynardie, Andreoni, Bellm, Bloom, Collett, Drake, Graham, Kasliwal, Kulkarni, Lemon, Miller, Neill, Nordin, Pierel, Richard, Riddle, Rigault, Rusholme, Sharma, Stein, Stewart, Townsend, Vinko, Wheeler, \& Wold}]{goobar_2023}
Goobar, A., Johansson, J., Schulze, S., {et~al.} 2023, Nature Astronomy, 7, 1098, \dodoi{10.1038/s41550-023-01981-3}

\bibitem[{{Grayling} {et~al.}(2026){Grayling}, {Thorp}, {Mandel}, {Pascale}, {Pierel}, {Hayes}, {Larison}, {Agrawal}, \& {Narayan}}]{Grayling2026MNRAS.548ag340G}
{Grayling}, M., {Thorp}, S., {Mandel}, K.~S., {et~al.} 2026, \mnras, 548, stag340, \dodoi{10.1093/mnras/stag340}

\bibitem[{{Hill} {et~al.}(2021){Hill}, {Lee}, {MacQueen}, {Kelz}, {Drory}, {Vattiat}, {Good}, {Ramsey}, {Kriel}, {Peterson}, {DePoy}, {Gebhardt}, {Marshall}, {Tuttle}, {Bauer}, {Chonis}, {Fabricius}, {Froning}, {H{\"a}user}, {Indahl}, {Jahn}, {Landriau}, {Leck}, {Montesano}, {Prochaska}, {Snigula}, {Zeimann}, {Bryant}, {Damm}, {Fowler}, {Janowiecki}, {Martin}, {Mrozinski}, {Odewahn}, {Rostopchin}, {Shetrone}, {Spencer}, {Mentuch Cooper}, {Armandroff}, {Bender}, {Dalton}, {Hopp}, {Komatsu}, {Nicklas}, {Ramsey}, {Roth}, {Schneider}, {Sneden}, \& {Steinmetz}}]{2021AJ....162..298H}
{Hill}, G.~J., {Lee}, H., {MacQueen}, P.~J., {et~al.} 2021, \aj, 162, 298, \dodoi{10.3847/1538-3881/ac2c02}

\bibitem[{{Horne}(1986)}]{Horne1986a}
{Horne}, K. 1986, \pasp, 98, 609, \dodoi{10.1086/131801}

\bibitem[{Hunter(2007)}]{hunter_matplotlib:_2007}
Hunter, J.~D. 2007, CSE, 9, 90, \dodoi{10.1109/MCSE.2007.55}

\bibitem[{{Johansson} \& {Goobar}(2026)}]{Johansson2026TNSAN.156....1J}
{Johansson}, J., \& {Goobar}, A. 2026, Transient Name Server AstroNote, 156, 1

\bibitem[{{Johansson} {et~al.}(2021){Johansson}, {Goobar}, {Price}, {Sagu{\'e}s Carracedo}, {Della Bruna}, {Nugent}, {Dhawan}, {M{\"o}rtsell}, {Papadogiannakis}, {Amanullah}, {Goldstein}, {Cenko}, {De}, {Dugas}, {Kasliwal}, {Kulkarni}, \& {Lunnan}}]{Johansson2021MNRAS.502..510J}
{Johansson}, J., {Goobar}, A., {Price}, S.~H., {et~al.} 2021, \mnras, 502, 510, \dodoi{10.1093/mnras/staa3829}

\bibitem[{{Johansson} {et~al.}(2025){Johansson}, {Perley}, {Goobar}, {Wise}, {Qin}, {McGrath}, {Schulze}, {Lemon}, {Gangopadhyay}, {Tsalapatas}, {Andreoni}, {Bellm}, {Bloom}, {Dekany}, {Dhawan}, {Fransson}, {Fremling}, {Graham}, {Groom}, {Gruen}, {Hall}, {Helou}, {Kasliwal}, {Laher}, {Lunnan}, {Mahabal}, {Miller}, {M{\"o}rtsell}, {Nordin}, {Hjortlund}, {Rich}, {Riddle}, {Singh}, {Sollerman}, {Townsend}, \& {Yan}}]{Johansson2025ApJ...995L..17J}
{Johansson}, J., {Perley}, D.~A., {Goobar}, A., {et~al.} 2025, \apjl, 995, L17, \dodoi{10.3847/2041-8213/ae1d61}

\bibitem[{Kelly {et~al.}(2015)Kelly, Rodney, Treu, Foley, Brammer, Schmidt, Zitrin, Sonnenfeld, Strolger, Graur, Filippenko, Jha, Riess, Bradac, Weiner, Scolnic, Malkan, Linden, Trenti, Hjorth, Gavazzi, Fontana, Merten, McCully, Jones, Postman, Dressler, Patel, Cenko, Graham, \& Tucker}]{kelly_2015}
Kelly, P.~L., Rodney, S.~A., Treu, T., {et~al.} 2015, Science, 347, 1123, \dodoi{10.1126/science.aaa3350}

\bibitem[{{Kelly} {et~al.}(2016){Kelly}, {Brammer}, {Selsing}, {Foley}, {Hjorth}, {Rodney}, {Christensen}, {Strolger}, {Filippenko}, {Treu}, {Steidel}, {Strom}, {Riess}, {Zitrin}, {Schmidt}, {Brada{\v{c}}}, {Jha}, {Graham}, {McCully}, {Graur}, {Weiner}, {Silverman}, \& {Taddia}}]{Kelly_2016}
{Kelly}, P.~L., {Brammer}, G., {Selsing}, J., {et~al.} 2016, \apj, 831, 205, \dodoi{10.3847/0004-637X/831/2/205}

\bibitem[{{Lemon} {et~al.}(2026){Lemon}, {Goobar}, {Johansson}, {M{\"o}rtsell}, {Schulze}, {Andreoni}, {Bochenek}, {Brennan}, {Busmann}, {Coughlin}, {Das}, {Dhawan}, {Fremling}, {Gangopadhyay}, {Gruen}, {Hall}, {Ho}, {Kasliwal}, {Perley}, {Rigault}, {Schroeder}, {Smith}, {Sollerman}, {Somalwar}, {Stein}, {Thorp}, {Townsend}, {Wise}, {Yan}, {Arendse}, {Bellm}, {Chen}, {Drake}, {Masci}, {Purdum}, {Smith}, {Hinkle}, {Rivera-Thorsen}, {Shappee}, {Tucker}, {Aguilar}, {Ahlen}, {Aldering}, {BenZvi}, {Bianchi}, {Brooks}, {Claybaugh}, {de la Macorra}, {Della Costa}, {Dey}, {Doel}, {Flaugher}, {Font-Ribera}, {Forero-Romero}, {Gazta{\~n}aga}, {Gontcho A. Gontcho}, {Gutierrez}, {Huterer}, {Ishak}, {Jimenez}, {Joyce}, {Juneau}, {Kehoe}, {Kim}, {Kirkby}, {Kisner}, {Kremin}, {Lahav}, {Landriau}, {Le Guillou}, {Levi}, {Manera}, {Meisner}, {Miquel}, {Moustakas}, {Nadathur}, {O'Connor}, {Palanque-Delabrouille}, {Palmese}, {Percival}, {P{\'e}rez-R{\`a}fols}, {Poppett}, {Prada}, {Rossi}, {Sanchez}, {Schlegel}, {Schubnell},
  {Shafieloo}, {Silber}, {Sprayberry}, {Tarl{\'e}}, {Weaver}, \& {Zou}}]{Lemon2026ApJ..1003L..47L}
{Lemon}, C., {Goobar}, A., {Johansson}, J., {et~al.} 2026, \apjl, 1003, L47, \dodoi{10.3847/2041-8213/ae6780}

\bibitem[{{Li} {et~al.}(2026){Li}, {Yan}, {Gkini}, \& {others}}]{Li2026}
{Li}, M., {Yan}, L., {Gkini}, A., \& {others}. 2026, ApJ, submitted

\bibitem[{{Marsh}(1989)}]{Marsh1989PASP..101.1032M}
{Marsh}, T.~R. 1989, \pasp, 101, 1032, \dodoi{10.1086/132570}

\bibitem[{{M{\"o}rtsell} {et~al.}(2026){M{\"o}rtsell}, {Johansson}, {Goobar}, \& {others}}]{Mortsell2026}
{M{\"o}rtsell}, E., {Johansson}, J., {Goobar}, A., \& {others}. 2026, ApJ, submitted

\bibitem[{{Oke} {et~al.}(1995){Oke}, {Cohen}, {Carr}, {Cromer}, {Dingizian}, {Harris}, {Labrecque}, {Lucinio}, {Schaal}, {Epps}, \& {Miller}}]{Oke95}
{Oke}, J.~B., {Cohen}, J.~G., {Carr}, M., {et~al.} 1995, \pasp, 107, 375, \dodoi{10.1086/133562}

\bibitem[{{Osman Hjortlund et al.}(2026)}]{Hjortlund2026}
{Osman Hjortlund et al.} 2026, in prep.

\bibitem[{{Pascale} {et~al.}(2025){Pascale}, {Frye}, {Pierel}, {Chen}, {Kelly}, {Cohen}, {Windhorst}, {Riess}, {Kamieneski}, {Diego}, {Meena}, {Cha}, {Oguri}, {Zitrin}, {Jee}, {Foo}, {Leimbach}, {Koekemoer}, {Conselice}, {Dai}, {Goobar}, {Siebert}, {Strolger}, \& {Willner}}]{Pascale2025ApJ...979...13P}
{Pascale}, M., {Frye}, B.~L., {Pierel}, J. D.~R., {et~al.} 2025, \apj, 979, 13, \dodoi{10.3847/1538-4357/ad9928}

\bibitem[{{Perley}(2019)}]{2019PASP..131h4503P}
{Perley}, D.~A. 2019, \pasp, 131, 084503, \dodoi{10.1088/1538-3873/ab2332}

\bibitem[{Pierel {et~al.}(2024)Pierel, Newman, Dhawan, Gu, Joshi, Li, Schuldt, Strolger, Suyu, Caminha, Cohen, Diego, DŚilva, Ertl, Frye, Granata, Grillo, Koekemoer, Li, Robotham, Summers, Treu, Windhorst, Zitrin, Agarwal, Agrawal, Arendse, Belli, Burns, Cañameras, Chakrabarti, Chen, Collett, Coulter, Ellis, Engesser, Foo, Fox, Gall, Garuda, Gezari, Gomez, Glazebrook, Hjorth, Huang, Jha, Kamieneski, Kelly, Larison, Moustakas, Pascale, Pérez-Fournon, Petrushevska, Poidevin, Rest, Shahbandeh, Shajib, Siebert, Storfer, Talbot, Wang, Wevers, \& Zenati}]{pierel_2024}
Pierel, J. D.~R., Newman, A.~B., Dhawan, S., {et~al.} 2024, The Astrophysical Journal Letters, 967, L37, \dodoi{10.3847/2041-8213/ad4648}

\bibitem[{{Pierel} {et~al.}(2024){Pierel}, {Frye}, {Pascale}, {Caminha}, {Chen}, {Dhawan}, {Gilman}, {Grayling}, {Huber}, {Kelly}, {Thorp}, {Arendse}, {Birrer}, {Bronikowski}, {Ca{\~n}ameras}, {Coe}, {Cohen}, {Conselice}, {Driver}, {D{\'S}ilva}, {Engesser}, {Foo}, {Gall}, {Garuda}, {Grillo}, {Grogin}, {Henderson}, {Hjorth}, {Jansen}, {Johansson}, {Kamieneski}, {Koekemoer}, {Larison}, {Marshall}, {Moustakas}, {Nonino}, {Ortiz}, {Petrushevska}, {Pirzkal}, {Robotham}, {Ryan}, {Schuldt}, {Strolger}, {Summers}, {Suyu}, {Treu}, {Willmer}, {Windhorst}, {Yan}, {Zitrin}, {Acebron}, {Chakrabarti}, {Coulter}, {Fox}, {Huang}, {Jha}, {Li}, {Mazzali}, {Meena}, {P{\'e}rez-Fournon}, {Poidevin}, {Rest}, \& {Riess}}]{Pierel2024ApJ...967...50P}
{Pierel}, J.~D.~R., {Frye}, B.~L., {Pascale}, M., {et~al.} 2024, \apj, 967, 50, \dodoi{10.3847/1538-4357/ad3c43}

\bibitem[{{Prochaska} {et~al.}(2020){Prochaska}, {Hennawi}, {Westfall}, {Cooke}, {Wang}, {Hsyu}, {Davies}, {Farina}, \& {Pelliccia}}]{prochaska_pypeit2020}
{Prochaska}, J., {Hennawi}, J., {Westfall}, K., {et~al.} 2020, The Journal of Open Source Software, 5, 2308, \dodoi{10.21105/joss.02308}

\bibitem[{{Qin et al.}(2026)}]{Qin2026}
{Qin et al.} 2026, in prep.

\bibitem[{{Ramsey} {et~al.}(1998){Ramsey}, {Adams}, {Barnes}, {Booth}, {Cornell}, {Fowler}, {Gaffney}, {Glaspey}, {Good}, {Hill}, {Kelton}, {Krabbendam}, {Long}, {MacQueen}, {Ray}, {Ricklefs}, {Sage}, {Sebring}, {Spiesman}, \& {Steiner}}]{1998SPIE.3352...34R}
{Ramsey}, L.~W., {Adams}, M.~T., {Barnes}, T.~G., {et~al.} 1998, in Society of Photo-Optical Instrumentation Engineers (SPIE) Conference Series, Vol. 3352, Advanced Technology Optical/IR Telescopes VI, ed. L.~M. {Stepp}, 34--42, \dodoi{10.1117/12.319287}

\bibitem[{Refsdal(1964)}]{refsdal_1964b}
Refsdal, S. 1964, Monthly Notices of the Royal Astronomical Society, 128, 307, \dodoi{10.1093/mnras/128.4.307}

\bibitem[{Rodney {et~al.}(2021)Rodney, Brammer, Pierel, Richard, Toft, O’Connor, Akhshik, \& Whitaker}]{rodney_2021}
Rodney, S.~A., Brammer, G.~B., Pierel, J. D.~R., {et~al.} 2021, Nature Astronomy, 5, 1118, \dodoi{10.1038/s41550-021-01450-9}

\bibitem[{{Shetrone} {et~al.}(2007){Shetrone}, {Cornell}, {Fowler}, {Gaffney}, {Laws}, {Mader}, {Mason}, {Odewahn}, {Roman}, {Rostopchin}, {Schneider}, {Umbarger}, \& {Westfall}}]{2007PASP..119..556S}
{Shetrone}, M., {Cornell}, M.~E., {Fowler}, J.~R., {et~al.} 2007, \pasp, 119, 556, \dodoi{10.1086/519291}

\bibitem[{{Stahl} {et~al.}(2020){Stahl}, {Mart{\'\i}nez-Palomera}, {Zheng}, {de Jaeger}, {Filippenko}, \& {Bloom}}]{Stahl_2020MNRAS.496.3553S}
{Stahl}, B.~E., {Mart{\'\i}nez-Palomera}, J., {Zheng}, W., {et~al.} 2020, \mnras, 496, 3553, \dodoi{10.1093/mnras/staa1706}

\bibitem[{{Storfer} {et~al.}(2026){Storfer}, {Wong}, {Acebron}, {Grillo}, {Hoogendam}, {Huang}, {Jones}, {Magnier}, {Mandel}, {Ratier-Werbin}, {Rubin}, {Shappee}, \& {Soler-Perez}}]{storfer2026arXiv260402418S}
{Storfer}, C.~J., {Wong}, K.~C., {Acebron}, A., {et~al.} 2026, arXiv e-prints, arXiv:2604.02418, \dodoi{10.48550/arXiv.2604.02418}

\bibitem[{{Suyu} {et~al.}(2024){Suyu}, {Goobar}, {Collett}, {More}, \& {Vernardos}}]{suyu2024SSRv..220...13S}
{Suyu}, S.~H., {Goobar}, A., {Collett}, T., {More}, A., \& {Vernardos}, G. 2024, \ssr, 220, 13, \dodoi{10.1007/s11214-024-01044-7}

\bibitem[{{Taubenberger} {et~al.}(2026){Taubenberger}, {Acebron}, {Ca{\~n}ameras}, {Chen}, {Galan}, {Grillo}, {Melo}, {Schuldt}, {Schweinfurth}, {Suyu}, {Aldering}, {Aryan}, {Lee}, {Mamuzic}, {Millon}, {Reynolds}, {Sergeyev}, {Asfandiyarov}, {Basa}, {Blondin}, {Burkhonov}, {Christensen}, {Courbin}, {Ehgamberdiev}, {Killestein}, {Mattila}, {Shaymanov}, {Shu}, {Xu}, {Yang}, {Gruen}, {Pierel}, {Storfer}, {Tran}, {Wong}, {Becerra}, {Dornic}, {Ducoin}, {Globus}, {Guti{\'e}rrez}, {Jiang}, {Kuncarayakti}, {L{\'o}pez-C{\'a}mara}, {Lundqvist}, {Magnani}, {M{\'e}ndez}, {Schneider}, \& {Vogl}}]{Taubenberger2026A&A...710A.365T}
{Taubenberger}, S., {Acebron}, A., {Ca{\~n}ameras}, R., {et~al.} 2026, \aap, 710, A365, \dodoi{10.1051/0004-6361/202557847}

\bibitem[{{Townsend} {et~al.}(2026){Townsend}, {Dhawan}, {Hayes}, \& {others}}]{Townsend2026}
{Townsend}, A., {Dhawan}, S., {Hayes}, E., \& {others}. 2026, ApJ, submitted

\bibitem[{{Turner} {et~al.}(2024){Turner}, {Smith}, \& {Collett}}]{turner2024MNRAS.528.3559T}
{Turner}, H.~C., {Smith}, R.~J., \& {Collett}, T.~E. 2024, \mnras, 528, 3559, \dodoi{10.1093/mnras/stae263}

\bibitem[{{Vazdekis} {et~al.}(2016){Vazdekis}, {Koleva}, {Ricciardelli}, {R{\"o}ck}, \& {Falc{\'o}n-Barroso}}]{Vazdekis2016MNRAS.463.3409V}
{Vazdekis}, A., {Koleva}, M., {Ricciardelli}, E., {R{\"o}ck}, B., \& {Falc{\'o}n-Barroso}, J. 2016, \mnras, 463, 3409, \dodoi{10.1093/mnras/stw2231}

\bibitem[{{Yan} {et~al.}(2018){Yan}, {Perley}, {De Cia}, {Quimby}, {Lunnan}, {Rubin}, \& {Brown}}]{Yan2018ApJ...858...91Y}
{Yan}, L., {Perley}, D.~A., {De Cia}, A., {et~al.} 2018, \apj, 858, 91, \dodoi{10.3847/1538-4357/aabad5}

\end{thebibliography}
